\pdfoutput=1
\documentclass[a4paper,12pt]{article}
\usepackage[utf8]{inputenc}
\usepackage[english]{babel}
\usepackage{amssymb,amsmath,amsfonts,amsthm}
\usepackage{graphicx,dsfont,color,hyphenat,subfig,setspace,caption,braket,bbold}
\usepackage[left=2.4cm,top=3.3cm,right=2.4cm,bottom=3.3cm,bindingoffset=0cm]{geometry}
\usepackage[titletoc,page]{appendix}
\usepackage[square,numbers,sort&compress]{natbib}
\usepackage[colorlinks=true,allcolors=blue]{hyperref}

\usepackage[normalem]{ulem}

\numberwithin{equation}{section}

\theoremstyle{definition}

\newcommand{\beq}{\begin{eqnarray}}
\newcommand{\eeq}{\end{eqnarray}}
\newcommand{\bea}{\begin{eqnarray}}
\newcommand{\eea}{\end{eqnarray}}
\newcommand{\be}{\begin{equation}}
\newcommand{\ee}{\end{equation}}
\newcommand{\non}{\nonumber\\}
\DeclareMathOperator{\Tr}{Tr}

\DeclareMathOperator{\U}{U}
\DeclareMathOperator{\SU}{SU}
\DeclareMathOperator{\SO}{SO}

\DeclareMathOperator{\diag}{diag}
\DeclareMathOperator{\MeV}{MeV}
\DeclareMathOperator{\GeV}{GeV}
\DeclareMathOperator{\fm}{fm}
\newcommand{\p}{\partial}
\renewcommand{\i}{\mathrm{i}}
\renewcommand{\d}{\mathop{}\!\mathrm{d}}

\newcommand{\Lag}{\mathcal{L}}
\newcommand{\bchi}{\boldsymbol{\chi}}
\newcommand{\btau}{\boldsymbol{\tau}}
\newcommand{\bpi}{\boldsymbol{\pi}}
\newcommand{\bx}{\mathbf{x}}
\newcommand{\bX}{\mathbf{X}}
\newcommand{\dM}{\delta{\mkern-2mu}M}
\newcommand{\calI}{\mathcal{I}}

\def\de{\partial}

\def\1{\mathbbm{1}}

\def\nn{\nonumber}

\def\xr{\xi^2+\rho^2}
\def\xx{\xi^2}
\def\rr{\rho^2}

\def\A{a}
\def\I{{\cal I}}

\newcommand{\hA}[0]{\widehat{A}}
\newcommand{\hF}[0]{\widehat{F}}

\numberwithin{equation}{section}

\begin{document}
\title{
\vskip 0pt
\bf{ \Large Mass and Isospin Breaking Effects \\ in the Skyrme Model and in Holographic QCD}
\vskip 15pt}
\author{
 Lorenzo Bartolini$^{(1)}$,  Stefano Bolognesi$^{(2)}$\ ,\\[-5pt]
     Sven Bjarke Gudnason$^{(1)}$
     \ and  Tommaso Rainaldi$^{(3)}$ \\[15pt]
{\em \footnotesize
$^{(1)}$Institute of Contemporary Mathematics, School of Mathematics and Statistics,
}\\[-10pt]
{\em \footnotesize
 Henan University, Kaifeng, Henan 475004, P.~R.~China
}\\[-10pt]
{\em \footnotesize
$^{(2)}$Department of Physics ``E. Fermi'', University of Pisa
  and INFN Sezione di Pisa
}\\[-10pt]
{\em \footnotesize
  Largo Pontecorvo, 3, Ed. C, 56127 Pisa, Italy}\\[-10pt]
{\em \footnotesize
  $^{(3)}$ Department of Physics, Old Dominion University, Norfolk, VA
  23529, USA  }
\\
{\footnotesize emails: lorenzo(at)henu.edu.cn, stefano.bolognesi(at)unipi.it,} \\[-10pt]
{\footnotesize gudnason(at)henu.edu.cn, train005(at)odu.edu
}}
\vskip 10pt
\date{December 2023}
\maketitle

\begin{abstract}
We discuss how the quark masses and their mass splitting affect the
baryons in the Skyrme model as well as the Witten-Sakai-Sugimoto (WSS)
model.
In both cases baryons are described by solitonic objects,
i.e.~Skyrmions and instantons, respectively.
After the quantization of their zeromodes the nucleons become quantum
states of a rotor.
We show how the quark mass affects the moment of inertia and we provide
a semi-analytic approach valid in the small-mass limit.
Additionally, we show how the two lightest quarks' mass splitting affects the
moments of inertia of the Skyrmion and induces an isospin breaking effect.
This effect turns out not to be enough to split the degeneracy in the
neutron-proton multiplet, but it splits some of the states in the
Delta multiplet.
Unlike the basic Skyrme model, the WSS model already includes vector mesons and another mechanism to
transfer isospin breaking from quark masses to the solitons is known.
We compute the splitting of the moment of inertia in the
small-mass limit in the WSS model and combine the two effects on the
spectrum of baryons, in particular the Deltas.
\end{abstract}

\newpage

\tableofcontents

\section{Introduction}

The Skyrme model \cite{Skyrme:1961vq,Skyrme:1962vh} provides us with a
tool to probe low-energy QCD and understand baryon phenomenology, see
the reviews \cite{Zahed:1986qz,Ma:2016npf}. 
Results can be obtained from a pure mesonic pseudo-scalar theory and
baryons are solitonic objects in this theory. The Skyrmion
configuration then needs to be quantized to obtain the full spectrum
of possible baryons: This can be done in the moduli space quantization
scheme \cite{Adkins:1983ya}.  
However, there is a missing piece of the bigger picture. 
One problem in particular, which is left unresolved in the
pseudo-scalar theory alone, is the isospin-breaking effect and how it
is transmitted from the quarks to the nucleons.

It is well known in the literature that the addition of the quark mass
in the Skyrme model induces a pion mass in the low-energy effective
theory \cite{Simic:1985gv}.
This has already been extensively studied within the Skyrme model
\cite{Adkins:1983hy,Battye:2004rw,Battye:2006tb,Houghton:2006ti,Battye:2006na}. 
%In this work, we are mostly interested in the
%effect of the (quark) mass on the moments of inertia of the Skyrmion, and
%consequently on the spectrum of the baryons.
The first part of this paper is devoted to the case of equal
quark masses and how this affects the moment of inertia of the Skyrmion.
We show that there is a way to treat the problem semi-analytically
in the limit of small mass, using the linear tail of the pions, where the
pion mass is induced by the quark mass.
In this regime the main correction to the moment of inertia
is linear in the pion mass $m$.
A nontrivial aspect of this computation is that the
first-order approximation has a divergence, which eventually is
resolved by the nonlinear core of the Skyrmion, but we are able to
prove that this term contributes only at order $m^2$ and thus can be
neglected at the linear order in $m$.
This contribution depends only on the coefficient in front of the
linear tail, which can be computed at $m=0$, and is related to the
$g_{NN\pi}$ coupling.
For the Skyrme model, we also perform the full computation
numerically; we have done it for the spherical case (quarks with equal
masses) and confirmed the validity of the above semi-analytic
approximation.
The physical intuition of why the dominant effect on the moment of
inertia from turning on a small pion mass comes from the tail of the
soliton is this. The small pion mass changes the solution only very
little in energy and hence the core of the solution is almost
unchanged. On the other hand, the tail of the soliton goes from a
power law at zero mass to an exponential falloff at a finite pion
mass, and hence the tail is the most sensitive to a tiny pion mass.
The total mass of the soliton changes only very little, but the moment
of inertia is roughly weighted with $r^2$ compared with the energy
density and hence is very sensitive to any change in the tail.

Splitting in baryonic masses is hard to obtain by first-principles
computations. The full theory of QCD should be tackled with
nonperturbative techniques, so the only viable options up to now have
been lattice formulation and chiral effective theories, like the Skyrme model.   
In this work, we will introduce isospin breaking in the quark mass
matrix
and examine how this affects the Skyrmion (baryon) configuration from the 
point of view of its symmetry properties, moduli space and quantum
states.

Adding an isospin-breaking term to the effective Lagrangian is not
difficult, and we want to do it in the most simple and natural way;
that is by introducing a splitting in the pion mass term. We know that
this is the right way to do it, as in the standard model the isospin
breaking is transferred from the Yukawa couplings to the quark mass
term and then to the pion mass term through the Gell-Mann-Oakes-Renner
(GMOR) relation. The first problem we encounter is that in the
$\SU(2)$ model alone, this term has no physical effect on the
pions. To make the effect visible, we have to extend the theory to
$\U(2)$ and thus consider also the $\eta$ meson.
We then have a measurable isospin splitting in the pion mass
matrix. But this is not the end of the story, the difficult part
then is to transfer this isospin breaking to the Skyrmion
sector. 
If we consider just the $\SU(2)$ pions and the quark mass splitting,
we do not see any effect at all, not even in the pion spectrum.
In order to see a splitting in the pion masses and consequently a
deformation of the Skyrmion, we need to introduce at least the $\eta$
field, the pseudo Nambu-Goldstone boson of the axial $\U(1)_A$
anomalous symmetry. With this minimally extended model, we can have 
isospin breaking in the pion masses and consequently a splitting in
the moments of inertia of the Skyrmion. 
  
The mass splitting of the pions affects the Skyrmion already at the
classical level, once we relax the assumption of the hedgehog
approximation. The Skyrmion remains axially symmetric, but becomes
prolate or oblate (this has to be determined and may be not obvious
\emph{a priori}). We devised a semi-analytic method to compute the
deformation of the inertia tensor in the small-mass limit. We then
tested this in the Skyrme model where the full numerical computations
can be made. 
Still the quantization of the Skyrmion moduli space does not provide a
mass splitting between neutron and proton states.
It provides a partial splitting in the $\Delta$ particles' masses.

It has been argued
\cite{Fujiwara:1984pk,Meissner:1988iv,Meissner:1986ka,Jain:1987sz,Schechter:1999hg,Kaymakcalan:1984bz,Zahed:1986qz}
that an extension of the Skyrme model with vector mesons can answer
many of the unsolved questions left from the pseudo-scalar theory.
Any precise model describing nucleon interactions should also take
into account heavier particles, such as the lightest vector mesons,
whose masses are around $780$ MeV; the presence of these new particles
certainly has an impact on the structure and the spectrum of
the Skyrmion, see e.g.~Refs.~\cite{Naya:2018mpt,Naya:2018kyi}.
Another reason, comes from the validity of the large-$N_c$ theory that
views baryons as solitons of an effective theory of both
pseudo-scalars and vector mesons that are actually predicted by the
large-$N_c$ theory \cite{Witten:1979kh}. 

The early literature on the proton-neutron mass splitting in the Skyrme
model context started with computations of a current-current
electromagnetic contribution to the splitting \cite{Durgut:1985mu} --
a formula derived in the bag model of quarks using the one-particle
exchange Feynman diagram \cite{Deshpande:1976vn}.
Their result was both incorrect in sign and numerically inaccurate, as
pointed out shortly after in Ref.~\cite{Ebrahim:1987mu}.
The intuition from the bag model was that the quark mass splitting in
the bag model would give rise to a too large positive mass splitting
between the neutron and the proton of about $1.79$ MeV, whereas the
subdominant negative current-current electromagnetic contribution
would be around $-0.50$ MeV, lowering the result of the mass splitting
to roughly the correct experimental value \cite{Ebrahim:1987mu}.
As was also pointed out in Ref.~\cite{Ebrahim:1987mu}, the $\SU(2)$
Skyrme model with isospin breaking in terms of a non-derivative
potential is not able to provide the contribution to the
neutron-proton mass splitting.
Soon after a new mechanism for introducing a non-electromagnetic
neutron-proton mass splitting in the Skyrme model was proposed in
Ref.~\cite{Epele:1988ak}, where experimental evidence of
non-electromagnetic isospin breaking in $\rho$-$\omega$ meson mixing
led the authors to introduce a term $\lambda\rho_\mu\omega^\mu$ into
the Skyrme model.
The neutron-proton mass splitting was not computed by computing the
impact on the Skyrmion itself, but rather by performing a
current-current computation of the $\rho$ meson and the $\omega$
meson, giving a formula of the same kind as the electromagnetic
contribution, albeit this time with the necessary positive sign
\cite{Epele:1988ak}.
The only role of the Skyrme model in this computation was to evaluate
the strong interaction form factors in this Feynman diagram of a
vector boson exchange between nucleons.
Using phenomenological input data, this result gave a splitting of the
right order of magnitude within the usual 30\% accuracy.

A result more in the spirit of using the Skyrmion as the nucleon, was
obtained by Jain-Johnson-Park-Schechter-Weigel 
\cite{Jain:1989kn}, where the model is extended from $\SU(2)$ to
$\U(2)$, by adding the $\eta$ meson and the $\rho$ and $\omega$ mesons
are added as well.
Their paper is concerned with getting the correct strong sector
contribution to the neutron-proton mass splitting and involves many
explicit chiral symmetry breaking terms of four categories as well as 
``cranking'' corrections, which are essential in their approach.
As for the mechanism of the neutron-proton mass splitting, their
result was nicely summarized in the review \cite{Schechter:1999hg},
where it was pointed out that the dominant source of the
neutron-proton mass splitting comes from the mass term that is
directly proportional to the down-up quark mass difference.
Although this term sources the $\eta$ field, the source is
proportional to the quark mass splitting $\epsilon$ making the
resulting field $\eta$ proportional to $\epsilon$ and in turn the
contribution to the energy proportional to $\epsilon^2$ and both the
neutron and the proton would receive the same correction.
Their mechanism works by including a source for the $\eta$ by an
approach dubbed a cranking mechanism, which consists of gauging an
anomaly term that contains the Wess-Zumino term and three other
anomalous pieces (see the term proportional to $d_1$ of Eq.~(3.11) in
Ref.~\cite{Jain:1987sz}).
The coefficient is determined by computing the electromagnetic
$\omega\to\gamma\pi^0$ decay rate \cite{Meissner:1988iv}.
The contribution of this ``gauging'' to the electroweak current acts as
a source for the $\eta$ that is not proportional to the quark mass
splitting, but requires the $\omega$ vector meson to the present.
This turning-on of the $\eta$ field then gives the dominant
contribution to the neutron-proton mass splitting from the pion mass
term, in the mechanism of Ref.~\cite{Jain:1989kn}. 

Results for the neutron-proton mass splitting were also obtained by
several groups in chiral bag models, which are conceptually somewhat
different from the Skyrme-type models, in that they include explicit
quark fields that become important in the core of the ``Skyrmion''
e.g.~\cite{Park:1989ak,Jain:1989kn}.
An obvious conceptual challenge with this approach, is whether there
is any over-counting in the nucleon of the quark degrees of freedom or
not, although this has been argued not to be a problem since the
quarks are dominantly present in the core of the nucleon and the pion
cloud is dominant outside the core.
We will not discuss the neutron-proton mass splitting in those models
here, but refer to the literature \cite{Park:1989ak,Jain:1989kn}.

It is worth mentioning that another recent model in the literature
\cite{Speight:2018zgc} achieved the neutron-proton mass 
splitting in the Skyrme model, by adding an isospin breaking term as a
derivative coupling between the $\omega$ meson and the charged pions.
Not including the neutral pion thus explicitly breaks isospin symmetry.
Although this mechanism is simple and provides a neutron-proton mass
splitting, it is not clear to us whether this term can be derived as
an effective Lagrangian term induced from the quark mass difference or
whether it can be embedded into a holographic model.
We want to pursue the more traditional approach of inducing only the
isospin breaking in the mass term, as this is what we expect from
chiral effective theory and the standard model.

Our result in Sec.~\ref{due} of this paper, is that we can
analytically compute the coefficient of the contribution to the moment
of inertia which is linear in the pion mass.
Since it is quadratic in the mass splitting parameter, it is also
quadratic in the isospin quantum number and provides no contribution
to the neutron-proton mass splitting, as mentioned above, but it
provides a splitting between the masses of the isospin-$\tfrac32$ and 
isospin-$\tfrac12$ Deltas.

In order to move towards the more realistic picture of the isospin
breaking effects, we thus need to take into account:
\begin{itemize}
\item The $\eta$ field (corresponding to extending $\SU(2)$ to $\U(2)$)
\item Providing a source of the $\eta$ that does not originate from
  the isospin breaking $\pi^0\eta$ vertex (the mass term itself).
\end{itemize}
Ref.~\cite{Jain:1989kn} used the $\omega$ meson and cranking to arrive
at a physical neutron-proton mass splitting.
We will choose a slightly different route, which we consider to be
somewhat easier than introducing the vector mesons in the Skyrme model
with the gauged WZW term, other gauged anomalous terms as well as the
$\eta$ field -- we will consider the Witten-Sakai-Sugimoto (WSS)
model, which contains all the vector mesons, the $\eta$ as well as the
gauged anomalous term as its 5-dimensional Chern-Simons term.
The advantage here, will be that we can utilize the
Belavin-Polyakov-Schwartz-Tyupkin (BPST) flat-space
instanton and we thus leave the similar further analysis of our paper
in the Skyrme model to the future.

In the WSS model \cite{Witten:1998qj,Sakai:2004cn,Sakai:2005yt}, the
properties of chiral symmetry are encoded in the gauge symmetry of the
fields living on the D8-branes.
The model shows spontaneous chiral symmetry breaking through the
merging of the antipodal stacks of D8/$\overline{\text{D8}}$-branes,
hence the symmetry group is reduced to
$\SU(N_f)_V\times\U(1)_A\times\U(1)_V$.
In the context of this holographic model, the pion matrix can be
obtained as the holonomy of the gauge field in the holographic
direction, and a baryon is realized as an instanton configuration of
the gauge field on the D8-branes, where the winding number of the
instanton has the natural interpretation as the baryon number. The
instanton on the boundary looks like a Skyrmion.
In this model, the vector mesons are automatically included; they come
from the gauge field fluctuations on the D8-branes.
A mechanism quite similar in spirit to that of Ref.~\cite{Jain:1989kn}
can be naturally embedded in the WSS model of holographic QCD
\cite{Bigazzi:2018cpg} and the neutron-proton mass splitting is in
fact much easier to compute in this setting. 

In the second part of this paper, we will look at two different mass
splitting effects: The newly found splitting of the moments of
inertia, which we computed in the Skyrme model and verified by full
numerical computations and the $\eta$-induced splitting that
contributes to the neutron-proton mass splitting.
In particular, we checked from the Skyrme model, that the correction
to the moment of inertia that is linear in the pion mass and
analytically calculable, is in fact the leading-order correction at
small pion masses.
Since the full numerical computation in the WSS model is much more
difficult, we trust the analytic result here due to the check in the
Skyrme model.
The lack of the full numerical computation in the WSS model means
e.g.~that we cannot accurately compute the Delta multiplet masses,
but we can quite accurately compute their splitting.
We will here consider the complete effective Hamiltonian that,
unlike the original scenario (\cite{Bigazzi:2018cpg}) produces a less
symmetric mass spectrum 
for baryons. The two effects of transfer of isospin breaking from
quarks to the baryon act in different ways; the first is linear and the
second is quadratic in $j_z$. 
When they are combined, we have a mass splitting between all four
states of the Delta multiplet as well as 
differences among the mass splittings.
The splitting between all four states in the Delta multiplet is an
$m^2$ effect, whereas a splitting between the isospin $3/2$ and $1/2$
states (with either sign) is a linear-in-$m$ effect.
In the limit of small $m$, and with our approximations, the rotor is
prolate.

The paper is organized as follows.
First we analyze the origin of the splittings in the Skyrme model in
Sec.~\ref{due} followed by a similar analysis for the holographic WSS
model in Sec.~\ref{tre}. 
We conclude with a discussion in Sec.~\ref{conclusion}.

%%%%%%%%%%%%%%%%%%%%%%%%%%%%%%%%%%%%%%%%%%%%%%%%%%%%%%%%%

\section{Skyrme model, mass and isospin breaking}
\label{due}

The Skyrme model is an effective Lagrangian whose degrees of freedom
are the pseudo-Goldstone bosons coming from the breaking of chiral
symmetry. The degrees of freedom are given compactly in the form of
the $\SU(N_f)$-valued field
$U(x)=e^{\i\phi(x)/f_\pi}$, which in the
two-flavor case, i.e.~$N_f=2$ reads
\begin{equation}
  U(x) = e^{\frac{\i\tau^a\pi^a(x)}{f_{\pi}}}\ ,
  \label{eq:U}
\end{equation}
where $\tau^a$, $a=1,2,3$, are the Pauli matrices and $f_{\pi}$ is the
pion decay constant.
Under chiral transformations $\SU(2)_L\otimes\SU(2)_R$ the field $U$ 
behaves as follows:
\begin{equation}
U\to V_LUV^{\dagger}_R\ ,\qquad
V_L\in\SU(2)_L,\qquad V_R\in\SU(2)_R \ .
\label{eq:U_field}
\end{equation}
The left-invariant current is
\begin{equation}
\begin{array}{l}
L_{\mu} = U^{\dagger}\partial_{\mu}U\to V_RL_{\mu}V^{\dagger}_R\ ,
\end{array}
\label{eq:left_right_currents}
\end{equation}
and thus transforms only under right-handed
transformations\footnote{There exists also a right-invariant current
$R_{\mu}\equiv\partial_{\mu}U U^{\dagger}\to V_LR_{\mu}V_L^{\dagger}$,
which transforms under left-handed transformations.
The Skyrme Lagrangian, however, can be equivalently expressed in terms
of either $L_{\mu}$ or $R_{\mu}$ and we choose the former.}.

In addition to the usual quadratic term ($\mathcal{L}_2$) of
the nonlinear sigma model (NLSM), the Skyrme model features a quartic
contribution in the derivatives known as the Skyrme term
($\mathcal{L}_4$), which is essential to the stability of the soliton,
as well as a mass term ($\mathcal{L}_0$), that explicitly breaks
chiral symmetry.
The complete effective theory Lagrangian thus reads: 
\begin{align}
  \Lag &= \Lag_2 + \Lag_4 + \Lag_0 \non
  &= \frac{f^2_{\pi}}{4}\Tr(L_{\mu}L^{\mu}) + \frac{1}{32e^2}\Tr\left([L_{\mu},L_{\nu}][L^{\mu},L^{\nu}]\right) + \frac{f^2_{\pi}c}{2}\Tr\left(M(U + U^{\dagger} -2\mathbb{1}_2)\right).
\label{eq:Skyrme_Lagrangian_massive}
\end{align}
Here $M\equiv$ diag$(m_u,m_{d})$ is the quark-mass matrix which, in
the two-flavored case, only contains the \textit{up} and \textit{down}
quark masses
\begin{equation}
\begin{split}
M &=  m_q\mathbb{1}_2 + \epsilon m_q\tau^3 \ ,\\
m_q &= \frac{1}{2}(m_u + m_d)\ , \qquad \epsilon = \frac{m_u - m_d}{m_u + m_d}\ .
\label{masscomplete}
\end{split}
\end{equation}
The parameter $c$ is defined such that, in the expansion of $\Lag_0$,
the mass term for the pions is correctly normalized, that is
\begin{equation}
\begin{split}
2cm_q&=c(m_u + m_d)\equiv m^2_{\pi^0} \simeq (139\MeV)^2\ ,\\
m_q&=3.45\MeV,\qquad
\epsilon = -0.34\ .
\end{split}
\end{equation}
By inspection of Eq.~\eqref{eq:Skyrme_Lagrangian_massive} we can tell
that no term proportional to $\epsilon$ appears at this stage.
This is because it would be proportional to the Pauli matrix $\tau^3$, whose
trace is identically zero and $U+U^\dag\propto\mathbb{1}_2$ for
$U\in\SU(2)$. (This would not be true for three flavors, see for
example \cite{Wu:1990ma,Kopeliovich:1997pq}.)

\subsection{The static Skyrmion}

The hedgehog Ansatz based upon the assumption of maximal symmetry is:
\begin{equation}
U(\bx) = e^{\i\btau\cdot\hat{\bx}f(r)} = \cos(f(r))\mathbb{1}_2 + \i(\btau\cdot\hat{\bx})\sin(f(r))\ ,
\label{eq:hedgehog_ansatz}
\end{equation} 
where $r\equiv |\bx|$, $\hat{\bx}=\bx/|\bx|$ and $f(r)$ is the profile function of the field. 
The \emph{spherical symmetry} implies that
any spatial rotation can be compensated  by a rotation in isospin space.
\begin{equation}
\begin{split}
U(R\bx) &= \cos(f(r))\mathbb{1}_2 + \i(\btau\cdot R\hat{\bx})\sin(f(r)) \\ & =
\cos(f(r))\mathbb{1}_2 + \i A(\btau\cdot\hat{\bx})A^{\dagger}\sin(f(r)) \ ,
\end{split}
\end{equation} 
where $R\in\SO(3)$, $A\in\SU(2)$ and the equality holds when
\begin{equation}
R_{ij} = \frac{1}{2}\Tr(\tau^iA\tau^jA^{\dagger}) \ .
\end{equation}

When the quark mass is zero the Skyrme model has two parameters
$f_{\pi}$ and $e$. Correspondingly the mass and the radius of the
Skyrmion $M_{\rm Sk}\sim \frac{f_{\pi}}{e}$,
$R_{\rm Sk} \sim\frac{1}{f_{\pi}e}$, where $\sim$ here means
``scales like''.
Energy and length rescaling can set these $f_\pi/e$ and $1/(f_\pi e)$
to unity and hence only the mass parameter changes the theory.
More precisely, only the dimensionless mass parameter
$\frac{m_{\pi^0}}{f_\pi e}$ (and the corresponding splitting parameter
$\frac{\epsilon m_{\pi^0}}{f_\pi e}$) change the theory and hence the
solutions.

After substituting the Ansatz \eqref{eq:hedgehog_ansatz} into
Eq.~\eqref{eq:Skyrme_Lagrangian_massive}, we obtain the following
expression for the Skyrmion mass
\begin{align}
  M^{\rm static}_{\rm Sk} 
  &=\frac{2\pi f_{\pi}}{e}\int\d r\,r^2\left[ f'^2 + \frac{2\sin^2f}{r^2}
    + \frac{\sin^2f}{r^2}\left(\frac{\sin^2f}{r^2} + 2f'^2\right)
    - 2m^2(\cos f-1) \right],
\label{eq:Skyrmion_mass_rescaled}
\end{align}
where we have rescaled the lengths as $r\to\frac{r}{f_{\pi}e}$ and
defined the dimensionless pion mass parameter
\beq
m \equiv \frac{m_{\pi^0}}{f_\pi e}\ .
\label{eq:pionmass_adim}
\eeq
The baryon number, which is given by the degree of the map
$U(\bx):S^3\mapsto SU(2)$, is given by
\beq
B = -\frac{2}{\pi}\int\d r\; \sin^2(f) f' \ .
\eeq
The equation of motion for the Skyrmion is found through the
minimization of its mass \eqref{eq:Skyrmion_mass_rescaled} and a
straightforward derivation yields: 
\begin{equation}
\left(1 + \frac{2\sin^2f}{r^2}\right)f'' + \frac{2f'}{r} + \frac{\sin 2f}{r^2}f'^2 -\frac{\sin 2f}{r^2} -\frac{\sin^2f\sin 2f}{r^4} - m^2\sin f = 0 \ .
\label{eq:EoM_spherical_massive}
\end{equation}
The solution of Eq.~\eqref{eq:EoM_spherical_massive} has to be found
within the functions that satisfy particular boundary conditions.
We already know that, at infinity, the $U$ field must be constant and
equal to the identity matrix $\mathbb{1}$, which implies
$f(r\to\infty)\to 0$. On the other hand, in order to provide a
Skyrmion with unit charge (i.e.~unit baryon number), the profile
function must also satisfy $f(0)=\pi$.
At large distances $f$ is small and the linearization of the equation
of motion \eqref{eq:EoM_spherical_massive} is
\beq
f'' + \frac{2f'}{r} - \frac{2f}{r^2} - m^2 f = 0\ ,
\eeq
and gives the exact asymptotic
behavior:
\begin{equation}
f(r) = C  \frac{  m r +1}{r^2} e^{-m r} \ ,\qquad r\gg 1 \ ,
\label{lineartailspheric}
\end{equation}
which in physical units (before rescaling the length scales)
corresponds to $r\gg (f_\pi e)^{-1}$. $C$ is a proportionality
constant in front of the linear tail and is related to the
nucleon-nucleon-pion coupling by \cite{Adkins:1983ya}:
\beq
g_{NN\pi} = \frac{8 \pi M_{\rm Sk}}{f_{\pi} e^2} C\ .
\eeq

\subsubsection{Numerical solutions}

\begin{figure}[!ht]
    \centering
    \includegraphics[width=0.4\textwidth]{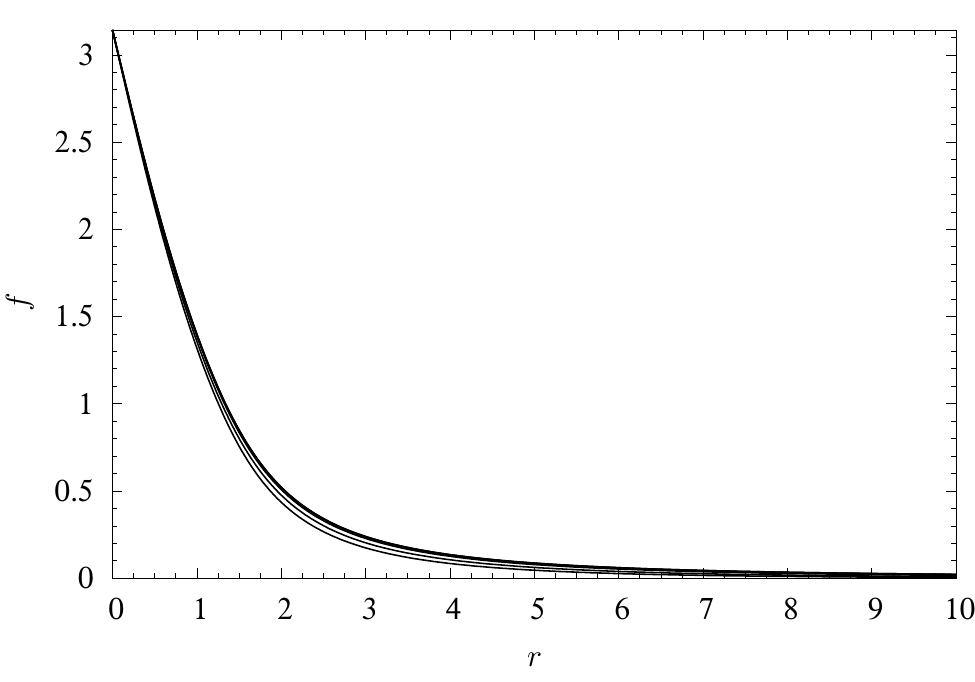}
    \includegraphics[width=0.4\textwidth]{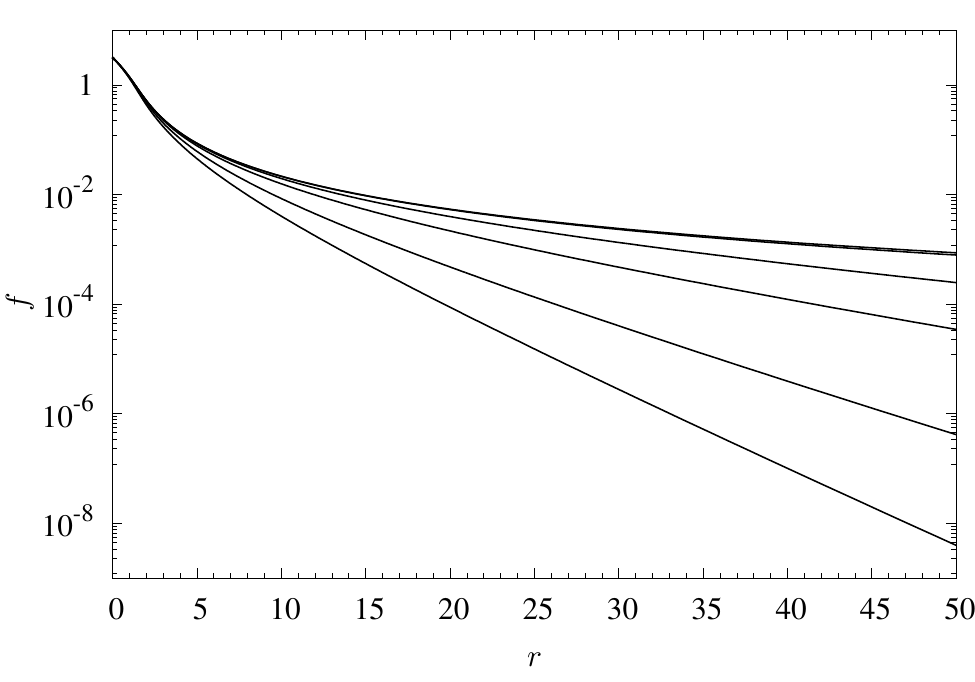}
    \caption{Profile function $f(r)$ for different values of the mass
      parameter $m=0,0.01, 0.05, 0.1, 0.2, 0.3, 0.5$, corresponding to
      the curves from top to bottom.
    }
    \label{profilemasssferic}
\end{figure}

We present some numerical solutions of
Eq.~\eqref{eq:EoM_spherical_massive} for various values of the
dimensionless mass parameter $m$ of Eq.~\eqref{eq:pionmass_adim}.
The profile function $f(r)$ is given in Fig.~\ref{profilemasssferic}
for values of the mass $m=0.01, 0.05, 0.1, 0.2, 0.3, 0.5$.
We will see shortly that the solutions need to be known very precisely
up to a very large distance. In particular, up to distances $r\gg1/m$,
which becomes a challenge for numerical methods in a box.
Because of the challenging numerical problem, we have decided to adopt
the following strategy. 
We use the shooting method from $r=r_{\rm min}\ll1$ with the
conditions $f(r_{\rm min})=\pi-f_pr_{\rm min}$ and $f'(r_{\rm min})=f_p$ and a second
shooting from $r=r_{\rm max}=50$ with $f$ and its derivative given by
Eq.~\eqref{lineartailspheric}.
The choice of $r=50$ is warranted since it is large enough that we can
trust the linearized solution (this will become apparent only a
posteriori), but not large enough to be the cutoff of the integrals,
in particular for the moment of inertia (see below). 
Next, we adjust $f_p$ and $C$ until the two solutions and their
derivatives match at a point in the middle taken to be
$r=r_{\rm mid}:=5$ (this point can be chosen arbitrarily).
Knowing the value of $C$ very precisely enables us to perform
integrals analytically from $r\in[r_{\rm max},\infty)$ using
the linearized solution \eqref{lineartailspheric}.

\begin{figure}[!ht]
    \centering
    \includegraphics[width=0.5\textwidth]{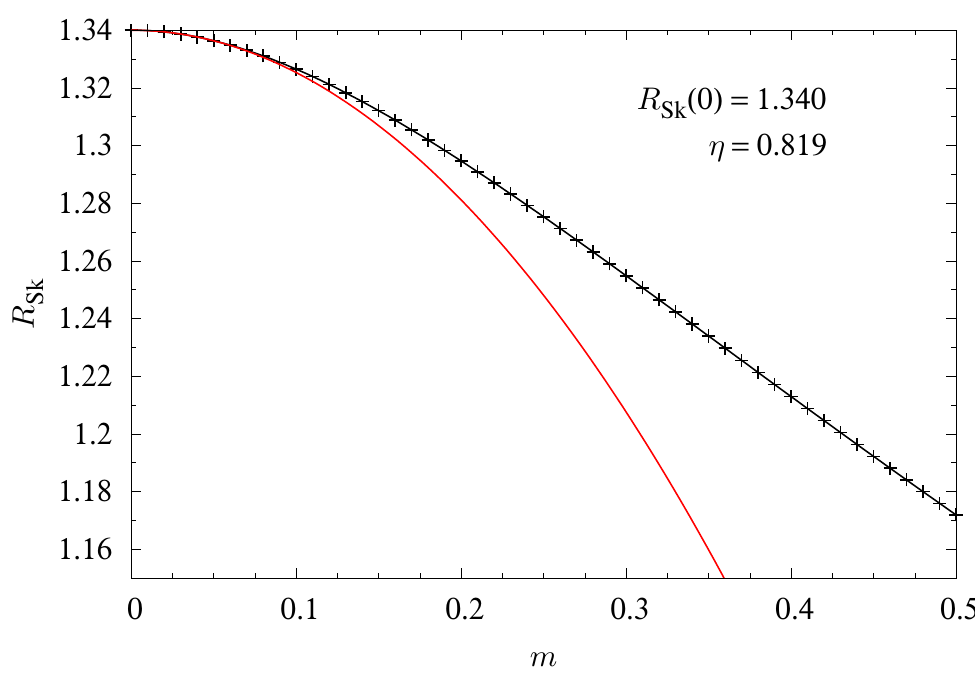}
    \caption{Skyrmion radius (defined by $f(R_{\rm Sk})=1$) as
      function of $m$.
      The red solid curve is a quadratic fit according to
      Eq.~\eqref{eq:Rm}.
    }
    \label{radiusmass}
\end{figure}

A way to understand and estimate the various parameters is to use 
Derrick's scaling argument \cite{Derrick:1964ww}, with which we can
show that the mass as function of the radius is given by
\beq
M(R) = \alpha_2 R + \frac{\alpha_4}{R} + \alpha_0m^2 R^3\ ,
\eeq
where the coefficients $\alpha_{0,2,4}$ come from the mass term, the
kinetic term and the Skyrme term, respectively, and can only be
determined numerically from solutions.
If we set $m=0$, minimization of $M(R)$ determines the leading-order
Skyrme radius $R_{\rm Sk}(0)=\sqrt{\alpha_4/\alpha_2}$.
Minimizing we get the Skyrmion radius
\beq
R_{\rm Sk}(m)^2 = R_{\rm Sk}(0)^2 \frac{\sqrt{1+8\eta m^2R_{\rm Sk}(0)^2}-1}{4\eta m^2 R_{\rm Sk}(0)^2}\ ,
\qquad
\eta = \frac{3\alpha_0}{2\alpha_2}\ .
\eeq
Expanding in small $m$, we obtain:
\beq
R_{\rm Sk}(m) \sim R_{\rm Sk}(0)\left(1 - \eta m^2 R_{\rm Sk}(0)^2 + \mathcal{O}(m^4)\right) \ .
\label{eq:Rm}
\eeq
We see that the mass term, to the smallest order, acts as a
compressing term shrinking the radius by an order $\sim m^2$. 
This is confirmed in Fig.~\ref{radiusmass}
where we defined the Skyrmion radius by 
\begin{equation}
f(R_{\rm Sk}) \simeq 1\ .
\end{equation}  
The soliton shrinks as the mass becomes larger. 

\begin{figure}[!ht]
    \centering
    \includegraphics[width=0.5\textwidth]{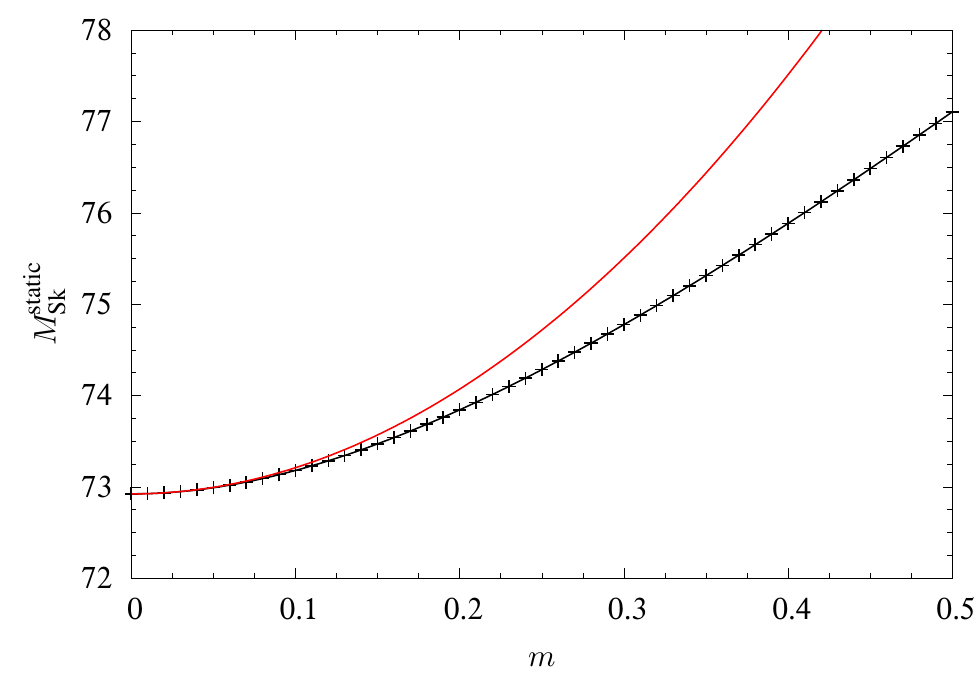}
    \caption{Skyrmion mass as function of $m$.
      The dots show the exact computation and the solid red curve is the
      quadratic approximation given in Eq.~\eqref{firstordermass}
      computed with the profile function for $m=0$.
    }
    \label{massmass}
\end{figure}

The mass of the Skyrmion is given in Fig.~\ref{massmass}. The mass is
increased by a quadratic term for small $m$ and it can be
evaluated semi-analytically. It is just the action term proportional to
$m^2$ evaluated on the solution obtained at $m=0$, which we denote as
$f_0$.
\begin{equation}
M^{\rm static}_{\rm Sk}(m) \simeq
M^{\rm static}_{\rm Sk}(0)
+  4\pi m^2\int\d r\;r^2\big[(1- \cos f_0)\big]\ .
\label{firstordermass}
\end{equation}
The validity of this approximation is confirmed by the results shown
in Fig.~\ref{massmass}, where the quadratic behavior characterizes the
small $m$ region of the curve.
This type of first-order correction is quite common, and found in many
examples \cite{Bolognesi:2014ova}.

\begin{figure}[!ht]
  \centering
  \mbox{
    \subfloat[]{\includegraphics[width=0.49\linewidth]{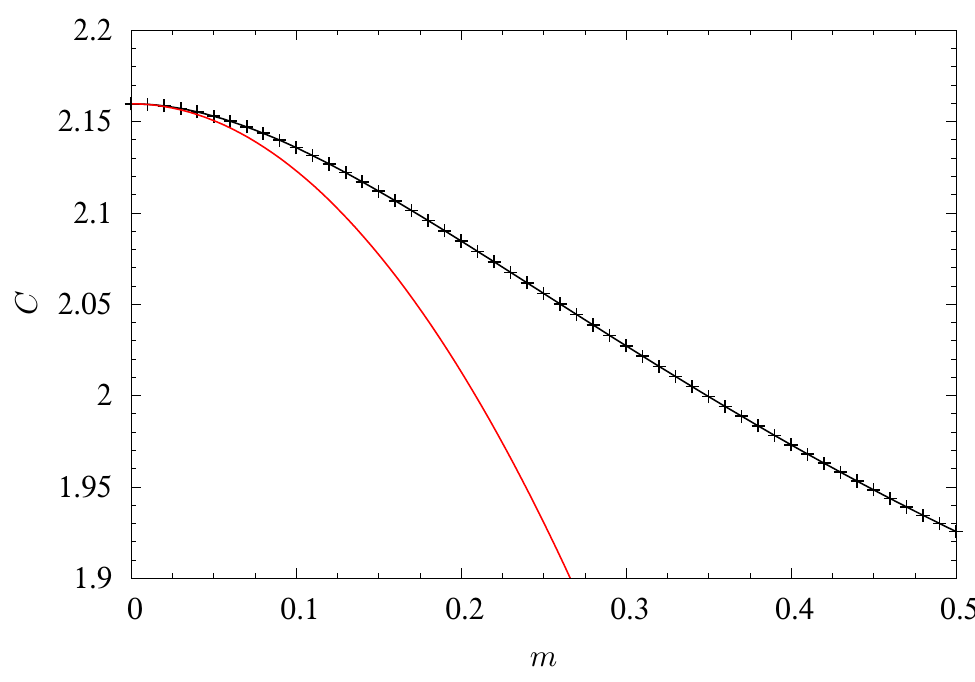}}
    \subfloat[]{\includegraphics[width=0.49\linewidth]{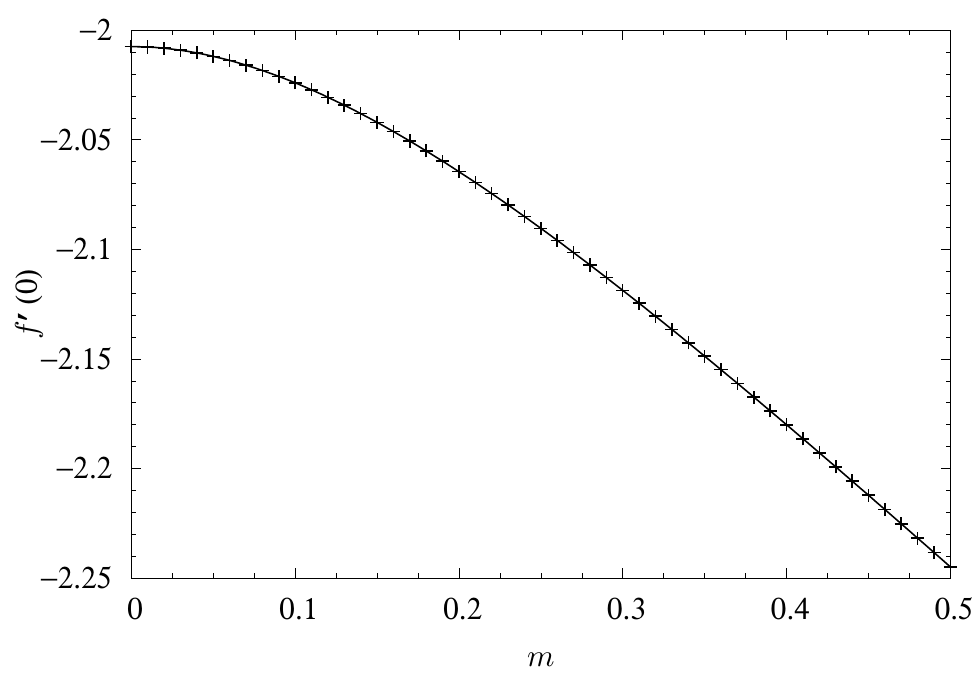}}}
    \caption{(a) Coefficient of the linear tail $C$ as function of $m$.
      The red curve is the approximation \eqref{eq:Cm} with $C(0)$
      fitted and $\eta$ from Fig.~\ref{radiusmass}.
      (b) The derivative of the profile function at the origin, $f'(0)$. 
    }
    \label{coeffmass}
\end{figure}

The coefficient $C(m)$ in front of the linear tail is shown in
Fig.~\ref{coeffmass}(a). A way to estimate $C(m)$ is the following. If we
assume that $f$ is dominated by the linear tail and we write the
condition $f(R_{\rm Sk})=1$, we have
\beq
 C(m)\frac{1 + m R_{\rm Sk}(m)}{R_{\rm Sk}(m)^2} e^{-m R_{\rm Sk}(m)} \simeq 1 \ .
\eeq
Solving for $C(m)$ and expanding to linear order, we obtain
\beq
C(m) = R(0)^2 
+ \frac{m^2}{2}\left(R(0)^4 + 2R(0)R''(0)\right)
+ \mathcal{O}(m^3) \ ,
\eeq
where $R(0)=R_{\rm Sk}(0)$, primes denote derivatives with respect to
$m$ and we have used that $R'(0)=\left.\frac{\d R}{\d m}\right|_{m=0}=0$, see 
Eq.~\eqref{eq:Rm}.
Inserting the latter equation gives us
\beq
C(m) = C(0) - 2m^2C(0)^2\left(\eta - \frac14\right)
+ \mathcal{O}(m^3)\ , \qquad
C(0) = R(0)^2\ .
\label{eq:Cm}
\eeq
From the fit shown in Fig.~\ref{radiusmass} we found that $\eta>1/4$.
This entails that $C(m)$, the coefficient of the linear tail,
decreases quadratically with $m$ and that is exactly what we see in
Fig.~\ref{coeffmass}(a), where the full numerical solution is
presented along with the quadratic fit for small $m$.

In Fig.~\ref{radiusmass}, the order $m^2$ correction is well predicted
by Eq.~\eqref{eq:Cm} using the computed value $R(0)=R_{\rm Sk}(0)$,
that is the coefficient of $m^2$ shown in the figure is $-2R(0)^4(\eta-1/4)$.
On the other hand, the prediction of the constant term $C(0)=R(0)^2$
turns out to be a bit coarse as $C(0)$ is computed to be $C(0)=2.160$,
whereas $R(0)^2=1.796$ and hence $17\%$ off.
Recall that we used rescaled lengths in the discussion above. It
will prove convenient to write down the value of the $C(0)$
coefficient in physical units for later comparison:
\beq
C(0) \simeq \frac{2.160}{(f_\pi e)^2}
\simeq 9.99\times 10^{-6} \MeV^{-2}
\simeq (0.624\fm)^2.
\eeq
In Fig.~\ref{coeffmass}(b) we show for completeness the other
``shooting parameter'' for the solution also as a function of $m$.
With the boundary conditions $f(0)=\pi$ and the two parameters $C$ and
$f'(0)$, the solution can be reconstructed fully.

We choose the physical values for the pion decay constant and the mass
of the uncharged pion, i.e.~$f_{\pi}=93\MeV$ and $m_{\pi^0}=139\MeV$;
the remaining parameter has been set to $e=5$ according to an estimate
given in Ref.~\cite{Ma:2016npf}, which is only slightly different with
the older fit, $e=5.45$ of Ref.~\cite{Adkins:1983ya}.
Using $e=5$ yields $m\simeq 0.3$ as value for the dimensionless pion
mass parameter.

Using numerical results for the pion profile function, we obtain the
rest mass of the soliton corresponding to both the massless and
massive cases (with $m=0.3$), see Table \ref{tab:rest_mass_estimates}.
\begin{table}[!htbp]
  \centering
  \begin{tabular}{ l | c | c }
    \hline
    \hline			
    $\,$  &$M^{\rm static}_{\rm Sk}$ (MeV) &  Final estimate (MeV)\\
    \hline
    Massless &$\approx \frac{f_{\pi}}{e}\times 72.92$   & $\approx 1356$\\
    Massive $(m=0.3)$ &$\approx \frac{f_{\pi}}{e}\times 74.78$  & $\approx 1391$\\
    \hline
    \hline  
  \end{tabular}
  \caption{Numerical results for the Skyrmion mass.}
  \label{tab:rest_mass_estimates}
\end{table}
Both results are greater than the average nucleon mass
$M_N\simeq 938.9$ MeV of about $44-48\%$.
We may ask if $m\simeq 0.3$ can be considered small; this depends on
the quantity we want to compute. If we consider the mass of the
Skyrmion, the first order approximation gives $72.92$ while the
correct numerical result is $74.78$; the approximation
\eqref{firstordermass} gives $75.50$, thus only $1.0\%$ from the exact
result.

\subsection{Quantization}

To describe baryons, we need to  excite the rotational degrees of
freedom of the Skyrmion. 
Following Ref.~\cite{Adkins:1983ya}, we introduce the $\SU(2)$
time-dependent collective variables $A(t)$ such that: 
\begin{equation}
 U(\bx,t) = A(t)U_0(\bx)A^{\dagger}(t)\ ,
\label{eq:U_time_A}
\end{equation}
where $U_0(\bx)$ is a classical static solution.
This corresponds to introducing a rigid isospin rotation, which when
quantized corresponds to the isospin quantum number (i.e.~the
difference between the number of protons and neutrons in a nucleus).
Due to the spherical symmetry of the hedgehog Ansatz, it is also
equivalent to rotation in configuration space by a matrix $R(t)$:
\begin{equation}
  U(\bx,t) = U_0(R(t)\bx)\ ,
\label{eq:U_time_R}
\end{equation}
which when quantized instead gives rise to the spin quantum number.
Obviously, for the spherically symmetric $B=1$ Skyrmion, the spin and
isospin quantum numbers must be equal in magnitude. 
Inserting Eq.~\eqref{eq:U_time_A} into
Eq.~\eqref{eq:Skyrme_Lagrangian_massive} we obtain:
\begin{equation}
\begin{split}
E_{\rm rot} &= 
\int \d^3x\,\left[-\frac{f^2_{\pi}}{4}\Tr(L_{0}L_{0}) - \frac{1}{16e^2}\Tr([L_{0},L_{i}][L_{0},L_{i}])\right] \\
&=  I\Tr(\dot{A}^{\dagger}\dot{A}) = \frac{1}{2}I\,\Omega^2\ ,
\label{eq:E_rot_p_E_static}
\end{split}
\end{equation}
where we have defined the angular velocity $\Omega$ so that:
\begin{equation}
\frac{\i}{2}\tau\cdot\Omega = -\dot{A}A^{\dagger}\ ,\qquad\textrm{or equivalently}\qquad
\Omega_i = -\frac{1}{2}\epsilon_{ijk}\big(\dot{R}R^{\rm T}\big)_{jk}\ .
\label{eq:angular_velocity_def}
\end{equation}
$I$ in Eq.~\eqref{eq:E_rot_p_E_static} is the moment of inertia of the
Skyrmion and it is an integral functional of the profile function:
\begin{equation}
  I = \int \d^3x\;\mathcal{I} =\left(\frac{1}{f_{\pi}e^3}\right)\frac{8\pi}{3}
  \int \d r\;r^2\sin^2f\Big( 1 + f'^2 + \frac{\sin^2f}{r^2} \Big)\ ,
\label{eq:inertia_moment_hedgehog}
\end{equation}
where we have first integrated out the angular dependence using that
\beq
\int\d\theta\d\varphi\;\sin\theta\,\hat{x}_i\hat{x}_j=
\frac{4\pi}{3}\delta_{ij}\ ,
\eeq
and rescaled $r\to \frac{r}{f_{\pi}e}$ as before.

\begin{figure}[!ht]
    \centering
    \includegraphics[width=0.5\textwidth]{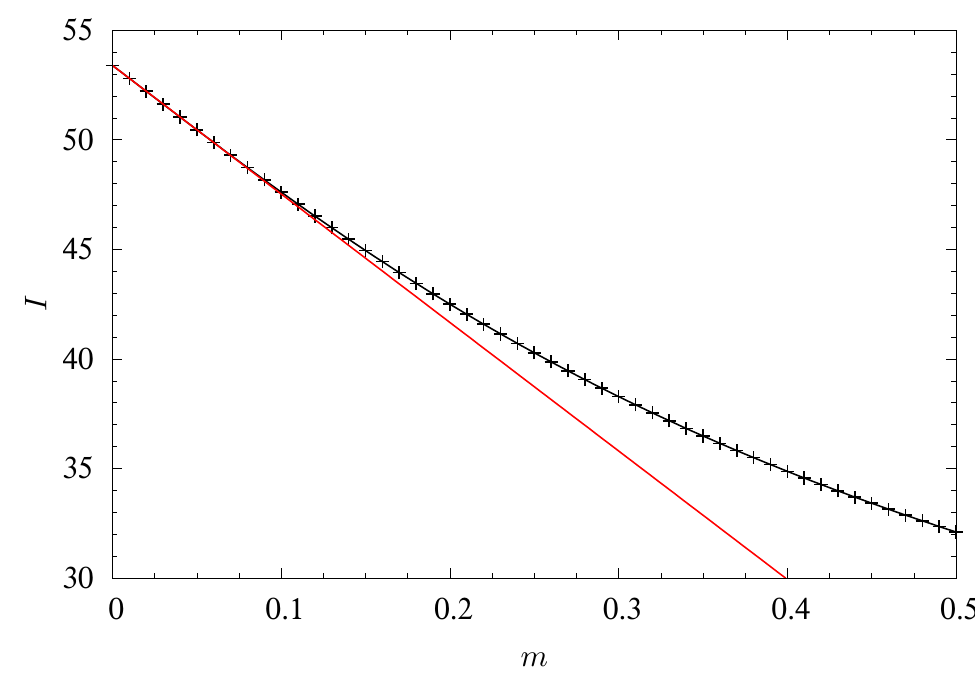}
    \caption{Moment of inertia for the Skyrmion as function of
      $m$. The slope of the red solid line computed analytically as
      $-4\pi C(0)^2$.
    }
    \label{inertiamass}
\end{figure}

Now we discuss the moment of inertia.
The exact function is \eqref{eq:inertia_moment_hedgehog}.
If we assume that $f$ is essentially given by the linear tail, we
get 
\begin{equation}
I  \simeq \frac{8\pi}{3}  \int_{R_{\rm UV}}^{\infty} \d r\;r^2 f^2\ ,\qquad
f(r) \simeq C(m)  \frac{1 + mr}{r^2} e^{-mr}  \ .
\label{linearI0}
\end{equation}
The above is only a lower bound on the moment of inertia since we are
completely discarding the contribution from the core of the soliton.
It is important to put a UV cutoff $R_{\rm UV}\simeq R_{\rm Sk}$, firstly
because the linear tail is a good approximation only up to the
Skyrmion core, secondly because this expression would be divergent
otherwise.
We can then perform the integral obtaining
\begin{equation}
  I \simeq \frac{8\pi}{3}  C(m)^2 e^{-2 m R_{\rm UV}}
  \left(\frac{m}{2} + \frac{1}{R_{\rm UV}}\right)\ .
\end{equation}
The UV divergence is the term $\sim \frac{1}{R_{\rm UV}}$ and when we
put $R_{\rm UV} \sim R_{\rm Sk}(m)$ we get exactly the order of
magnitude of $I(0)$. Expanding in powers of $m$ we get 
\begin{align}
  I(m) &= \frac{8\pi}{3}C(0)^2\left(\frac{1}{R_{\rm UV}} - \frac32m + \mathcal{O}(m^2)\right)\non
  &= I(0) - 4\pi C(0)^2 m + O(m^2)\ .
\label{linearI}
\end{align}
The coefficient in front of the linear term can be computed exactly,
and it is $-4\pi C(0)^2$.
This linear behavior with the exact coefficient is confirmed in
Fig.~\ref{inertiamass}.
It is important to stress the conditions that makes the linear
approximation \eqref{linearI} computable. There is an $m$ dependence
both in $C(m)$ and in $R_{\rm UV}\sim R_{\rm Sk}(m)$. This dependence
is quadratic as we saw before, so it does not affect the linear term
in $m$. So Eq.~\eqref{linearI} becomes exactly computable, apart from
$C(0)$ that has to be extracted from the numerical solution at $m=0$.

We compute the integral in Eq.~\eqref{eq:inertia_moment_hedgehog} for
massless and massive pions; 
the results are listed in Table \ref{tab:inertia_moment_estimates}.  
\begin{table}[!htbp]
  \centering
  \begin{tabular}{ l | c | c }
    \hline
    \hline			
    $\,$  &$I $ (MeV$^{-1}$) &  Final estimate (GeV$^{-1}$)\\
    \hline
    Massless &$\approx \frac{1}{f_{\pi}e^3}\times 53.38 $     & $\approx 4.592$\\
    Massive &$\approx \frac{1}{f_{\pi}e^3}\times 38.28 $     & $\approx 3.293$\\
    \hline
    \hline  
  \end{tabular}
  \caption{Numerical results for the moment of inertia of the Skyrmion.}
  \label{tab:inertia_moment_estimates}
\end{table} 
The first order correction linear in $m$ is a tail effect, or more
precisely due to the ``lack of tail''.
For $0\neq m\ll 1$, we can consider the solution to be roughly equal
to the $m=0$ solution, up to the scale $r_m \sim \frac{1}{m}$ where
everything falls off exponentially.
The moment of inertia on the $m=0$ solution is then reduced by the
fact that beyond $r_m$ the field is practically zero.
The lack of tail beyond $r_m$ is the reason for the negative linear
contribution in $m$ (Eq.~\eqref{linearI}).
This is quite different in nature than the first order contribution to
the mass (Eq.~\eqref{firstordermass}) which is instead due to the
unperturbed solution evaluated on the perturbed term, and thus is a
``core'' effect.
For a numerical evaluation of the moment of inertia at small mass, it
is very important to have a large IR cutoff, at least bigger than the
scale $r_m$. 
If we consider the inertia of the Skyrmion, the correct numerical
result is $38.28$, and the approximation \eqref{linearI} gives $35.80$,
thus only $6.5\%$ from the correct result; this not as close as what
happened for the mass, but we can still consider $m$ small for the
computation of $I$.

The next step is the quantization of the rotational energy. 
The conjugate classical momentum of the angular velocity $\Omega$ is:
\begin{equation*}
  J^i = \frac{\partial E_{\rm rot}}{\partial\Omega_i} = I\Omega^i \ ,
\end{equation*}
which means that the rotational energy can be re-expressed as:
\begin{equation*}
  E_{\rm rot} = \frac{J^2}{2I} \ .
\end{equation*}
Following the quantum mechanical rules for the quantization of the
angular momentum, we have that the eigenvalues of the total energy or
Skyrmion mass are: 
\begin{equation}
  M_{\rm Sk} = M^{\rm static}_{\rm Sk}
  + \frac{1}{2I}j(j+1)\ ,\qquad
  j = 0,\frac{1}{2},1,\frac{3}{2}, \ldots
  \label{eq:tot_hamiltonian_Skyrmion}
\end{equation}
where we have set $\hbar = 1$.
The rigid rotor quantization is valid as long as the energy splitting
of the rotational states are much smaller than any vibrational states
that deform the rotor.
From a theoretical point of view it is always possible to achieve this
limit by sending $\hbar\to0$. Here we will always work in this
approximation.
Changing to the energy units $\frac{f_\pi}{e}$ and length units
$(f_\pi e)^{-1}$, the would-be $\hbar=e^{-2}$.
From a phenomenological point of view we have $e\simeq 5$, so
$e^{-2}\simeq 0.04$ which can be considered small.

We can explicitly construct the physical quantities and 
quantum operators in terms of the   parametrization of the
$\SU(2)$-valued $A(t)$ matrix as \cite{Adkins:1983ya}:
\begin{equation}
  A(t) = a_0\mathbb{1} + \i a_k\tau^k\ ,\qquad
  a_0,a_k\in\mathbb{R}\ ,\qquad\text{and}\qquad a_0^2 + a_k^2 = 1\ .
\end{equation}
The Lagrangian becomes a function of the $a$'s and its temporal
derivatives, i.e.~$\Lag=\Lag(a,\dot{a})$.  
Using Eq.~\eqref{eq:angular_velocity_def}, the angular velocity $\Omega$ 
in terms of the local parameters is: 
\begin{equation}
  \Omega_i(a,\dot{a}) = 2(a_i\dot{a}_0 - a_0\dot{a}_i +
  \epsilon_{ijk}a_j\dot{a}_k)\ ,
\end{equation}
where the nonlinear constraint $a_0^2+a_k^2=1$ has been used to make
$\dot{A}A^\dag$ and hence $\tau\cdot\Omega$ $\mathfrak{su}(2)$-valued,
and the conjugate momentum to $a_{\mu}$ is: 
\begin{equation}
  p_{\mu} = \frac{\partial L}{\partial\dot{a}_{\mu}} =
  4I\dot{a}_{\mu}\ ,\qquad
  \mu = 0,1,2,3\ ,
  \label{eq:conjugate_momentum_a}
\end{equation}
where $p_\mu^2$ is the Laplacian on the 3-sphere
\cite{Adkins:1983ya}.
Notice that $\mu$ is here not a spacetime index, but is an
index on the Euclidean 3-sphere and therefore we do not distinguish
lower and upper indices.
We have all we need to write down the Hamiltonian associated to the
rotational modes of the Skyrmion, i.e.:
\begin{equation}
  H = p_{\mu}\dot{a}_{\mu} - L
  = M^{\rm static}_{\rm Sk} + \frac{1}{8I}p_\mu^2\ .
\end{equation}

\begin{figure}[!ht]
    \centering
    \includegraphics[width=0.7\textwidth]{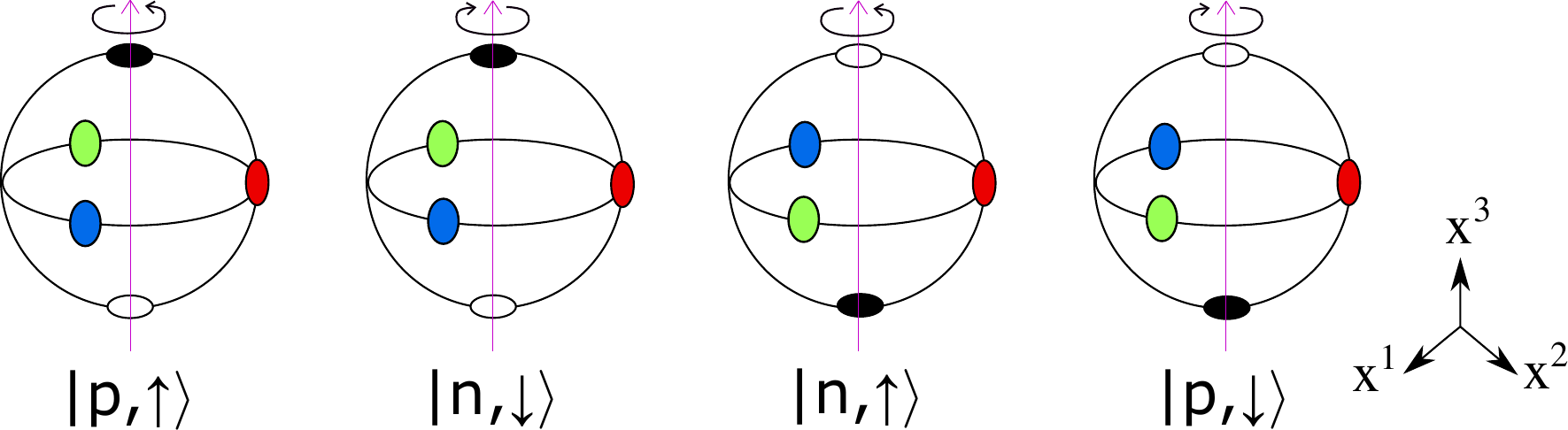}
    \caption{Schematic representations of the four states
      corresponding to the fundamental representation of $\SU(2)$. The
      proton spins in both physical and isospin (colors) space in the
      same direction, whereas the neutron behaves oppositely.
      The colors are representing the ``direction'' of the pions
      wrapping the $\SU(2)$ target space, with
      $\hat\pi_1+\i\hat\pi_2=e^{\i\theta}$ and $\theta=0,2\pi/3,4\pi/3$
      corresponding to red, green blue, respectively,
      where
      $\hat\pi^a=\frac{\pi^a}{\sqrt{\pi^b\pi^b}}$ is the normalized
      pion field of Eq.~\eqref{eq:U}.
      White and black correspond instead to the $\hat\pi_3=\pm1$,
      respectively.
      The ``classical'' version of the spinning nucleons has been discussed in Ref.~\cite{Foster:2015cpa}.
    }
    \label{fig:4_states}
\end{figure}

We finally notice that the rotational Lagrangian in
Eq.~\eqref{eq:E_rot_p_E_static} is manifestly invariant under two
types of $\SU(2)$ global transformations, ``rotations'' and
``isorotations'' respectively, namely: 
\begin{equation*}
  A(t)\to A(t)B\qquad\text{and}\qquad
  A(t)\to BA(t)\ ,\qquad\text{with}\qquad
  B\in\SU(2)\ .
\end{equation*}
The classical charges associated to these transformations can easily
be found and they represent the Skyrmion's spin and isospin:
\begin{equation}
  J_i = \i I\Tr\big(\tau^i\dot{A}A^\dag\big)\ ,\qquad
  I_i = -\i I\Tr\big(\tau^i A^\dag\dot{A}\big)\ , \qquad A = A(t)\ ,
\end{equation}
or in their quantum operator forms, after using
Eq.~\eqref{eq:conjugate_momentum_a} and making the replacement
$p_\mu\to -\i\frac{\partial}{\partial a_{\mu}}$:
\beq
  J_i = -\frac{\i}{2}\left(a_0\frac{\partial}{\partial a_i} - a_i\frac{\partial}{\partial a_0} + \epsilon_{ijk}a_j\frac{\partial}{\partial a_k}\right),\\
  I_i = -\frac{\i}{2}\left(a_i\frac{\partial}{\partial a_0}-a_0\frac{\partial}{\partial a_i} + \epsilon_{ijk}a_j\frac{\partial}{\partial a_k}\right).
\eeq
We can now classify the matrix element of $A(t)$ in terms of their
spin and isospin eigenvalues so to determine the neutron and proton
states \cite{Adkins:1983ya}, as shown in
Tab.~\ref{tab:neutron_proton_states}, see also Fig.~\ref{fig:4_states}.
\begin{table}[!htbp]
  \centering
  \begin{tabular}{ l | c | c | c |r }
    \hline
    \hline			
    $\,$  &$A_{11}$ & $A_{12}$ & $A_{21}$ & $A_{22}$ \\
    \hline
    $J_3$ &$1/2$     & $-1/2$   & $1/2$    & $-1/2$ \\
    $I_3$ &$-1/2$     & $-1/2$   & $1/2$    & $1/2$ \\
    $\,$  & $\ket{n\uparrow}$ &$\ket{n\downarrow}$ & $\ket{p\uparrow}$ &$\ket{p\downarrow}$\\
    \hline
    \hline  
  \end{tabular}
  \caption{Matrix elements of $A(t)$ corresponding to the fundamental
    spinor representation.} 
  \label{tab:neutron_proton_states}
\end{table}
In particular, it can be verified that higher representations can be
constructed from monomials in elements of $A(t)$, that is: Let
$A(t)_{ij}^l$ be the $l$-\emph{th} power of any matrix element of $A(t)$,
then it carries $J=I=l/2$, see Tab.~\ref{tab:delta_states}.
\begin{table}[!htbp]
  \centering
  \begin{tabular}{ l | c | c | c | c | c | c | c | c }
    \hline
    \hline			
          &$A_{11}^3$ & $A_{11}^2A_{12}$ & $A_{11}^2A_{21}$ & $A_{11}^2A_{22}$ & $A_{11}A_{21}^2$ & $A_{12}A_{21}^2$ & $A_{21}^2A_{22}$ & $A_{21}^3$ \\
          &          &                &                & $A_{11}A_{12}A_{21}$              & $A_{11}A_{21}A_{22}$ & & \\
    \hline
    $J_3$ &$3/2$    & $1/2$  & $3/2$  & $1/2$  & $3/2$ & $1/2$ & $1/2$    & $3/2$ \\
    $I_3$ &$-3/2$   & $-3/2$ & $-1/2$ & $-1/2$ & $1/2$ & $1/2$ & $3/2$    & $3/2$ \\
          & $\ket{\Delta_-\!\uparrow\uparrow}$ &$\ket{\Delta_-\!\uparrow}$ & $\ket{\Delta_0\!\uparrow\uparrow}$ & $\ket{\Delta_0\!\uparrow}$ & $\ket{\Delta_+\!\uparrow\uparrow}$ & $\ket{\Delta_+\!\uparrow}$ & $\ket{\Delta_{++}\!\uparrow}$ &$\ket{\Delta_{++}\!\uparrow\uparrow}$\\
    \hline
    \hline  
  \end{tabular}
  \caption[Short Heading]{Matrix elements of $A(t)$ corresponding to
    the $\Delta$ representation. Only the positive spins are displayed
  in this table. }
  \label{tab:delta_states}
\end{table}

In nature baryons carrying half-integer spin and isospin equal to
$1/2$ and $3/2$ are nucleons ($n$ and $p$) and $\Delta$ resonances,
respectively. From Eq.~\eqref{eq:tot_hamiltonian_Skyrmion}, we can
actually give an estimate for their mass, i.e.: \cite{Adkins:1983ya}
\begin{equation}
\begin{split}
&M_{n,p} = M^{\rm static}_{\rm Sk} + \frac{3}{8I}\ , \quad \quad M_{\Delta} = M^{\rm static}_{\rm Sk} + \frac{15}{8I}\  ,\\ 
& \qquad \qquad  \qquad  \ M_{\Delta}-M_{n,p} = \frac{3}{2I}\ ,
\end{split}
\label{eq:N_mass_and_Delta_mass}
\end{equation}
with the numerical results summarized in
Tab.~\ref{tab:final_mass_estimates}.   
The model we have studied does still not account for the mass
difference between the neutron and the proton, nor for the mass
difference between the $\Delta$'s. 
This is due to the fact that no isospin breaking term appears in the
Hamiltonian \eqref{eq:tot_hamiltonian_Skyrmion}.
\begin{table}[!htbp]
  \centering
	\begin{tabular}{ l | c | c | c}
		\hline
		\hline			
		$\,$  &$M_{n,p}$ (MeV) &  $M_{\Delta}$ (MeV)  &$\frac{M_{\Delta}-M_{n,p}}{M_{n,p}}$ \\
		\hline
		Massless &$\approx 1438$     & $\approx 1765$ &  $\approx 0.23$   \\
		Massive  &$\approx 1505$     & $\approx 1960$ &  $\approx 0.30$    \\  
                Real &$\approx  939 $     & $\approx 1232 $ &  $\approx 0.31 $   \\
		\hline
		\hline  
	\end{tabular}
	\caption{Results for the nucleon and $\Delta$ masses. Here the
        $f_\pi=93\MeV$, $e=5$ and the massive case has $m\simeq0.3$.}
	\label{tab:final_mass_estimates}
\end{table}

\subsection{Adding the \texorpdfstring{$\eta$}{eta} contribution}

The first step to modify the old theory is by enlarging the symmetry
group from $\SU(2)$ to $\U(2)$, that is, also considering the $\U(1)$
generator whose corresponding field we shall call $\eta$: 
\begin{equation}
  U = e^{\frac{\i\pi^a\tau^a}{f_{\pi}}}\rightarrow U'
  = U e^{\frac{\i\eta}{f_\pi}} = e^{\frac{\i(\pi^a\tau^a + \eta\mathbb{1}_2)}{f_{\pi}}}\ ,
\end{equation}
where we have set the singlet decay constant equal to $f_{\pi}$, since
we are working in the large-$N_c$ expansion setup. 
The $\eta$ field corresponds to the generator of the unbroken
$\U(1)_A$ symmetry group, which produces the axial anomaly.
Even in the chiral limit this particle shows a nonvanishing mass,
which is directly related to the axial anomaly.
The insertion of a phase factor generated by the $\eta$ field changes
the left-invariant current in the following way: 
\begin{equation}
L_{\mu}(\pi)\rightarrow L_{\mu}(\pi) +\frac{\i}{f_{\pi}}\partial_{\mu}\eta\, \mathbb{1}_2 \ .
\label{eq:L_eta}
\end{equation} 
The Skyrme term $\Lag_4$ is left unchanged and no $\eta$ dependence
can be present since it only appears as an additional term
proportional to the identity matrix.  
Although the Skyrme term does not change, the other two terms,
$\Lag_2$ and $\Lag_0$ are modified.
The new Lagrangian is written in the same way as before.  
We have chosen not to include the anomaly-related mass of $\eta$
since, in the limit where $N_c\to\infty$, it vanishes as it is of
order $1/N_c$.
However, this might not be the case in the real world where $N_c=3$
and a more detailed and careful analysis may be necessary.
We can thus write down the static Lagrangian as
\begin{align}
  \Lag &= -\frac12\partial_\mu\eta\partial^\mu\eta
  +\frac{f^2_{\pi}}{4}\Tr(L_{\mu}L^{\mu})
  + \frac{1}{32e^2}\Tr\left([L_{\mu},L_{\nu}][L^{\mu},L^{\nu}]\right)
  + f_\pi^2m_{\pi^0}^2\left(\sigma\cos\frac{\eta}{f_\pi} - 1\right)\non
  &\phantom{=\ }
  - \epsilon f_\pi^2m_{\pi^0}^2\bpi^3\sin\frac{\eta}{f_\pi}\ ,
  \label{eq:Skyrme_Lagrangian_eta}
\end{align}
where we keep $L_\mu$ as the left-invariant current of the $\SU(2)$
matrix $U$ and we have conveniently defined
\beq
U = e^{\frac{i\pi^a\tau^a}{f_{\pi}}}
= \sigma\mathbb{1}_2 + \i\btau\cdot\bpi
= \phi^0\mathbb{1}_2 + \i\tau^a\phi^a\ .
\eeq
Notice that the relation between $\pi^a$ and $\bpi^a$ is
\beq
\bpi^a=\frac{\pi^a}{\sqrt{\pi^b\pi^b}}\sin\left(\frac{\sqrt{\pi^c\pi^c}}{f_\pi}\right)\ .
\eeq
The mass term now shows an $\epsilon$ dependence, strictly related to
the ``existence'' of $\eta$.
We note that a spherical solution using the Ansatz
\eqref{eq:hedgehog_ansatz}, for which both $\bpi$ and $\eta$ are just
functions of the radius $r$, cannot yield the newfound breaking term
proportional to the quark mass splitting, $\epsilon$: 
\begin{align}
  \epsilon f_\pi^2m_{\pi^0}^2\int\d^3x\;\bpi^3\sin\frac{\eta}{f_\pi}
  &=2\pi\epsilon f_\pi^2m_{\pi^0}^2\int \d r\d\theta\; r^2\sin\theta
  \cos\theta\sin f(r)\sin\left(\frac{\eta(r)}{f_\pi}\right)\non
  &\propto\int_0^\pi\d\theta\sin(2\theta) = 0\ .
\end{align}
We now perform the rescaling $x^\mu\to\frac{x^\mu}{f_\pi e}$ and
$\eta\to\eta f_\pi$, obtaining:
\begin{align}
  \frac{\Lag}{f_\pi^4e^2} &=
  -\frac12\p_\mu\eta\p^\mu\eta
  -\frac12\p_\mu\sigma\p^\mu\sigma
  -\frac12\p_\mu\bpi\cdot\p^\mu\bpi\non
  &\phantom{=\ }
  -\frac14\left(\p_\mu\sigma\p^\mu\sigma + \p_\mu\bpi\cdot\p^\mu\bpi\right)^2
  +\frac14\left(\p_\mu\sigma\p_\nu\sigma + \p_\mu\bpi\cdot\p_\nu\bpi\right)\left(\p^\mu\sigma\p^\nu\sigma + \p^\mu\bpi\cdot\p^\nu\bpi\right)\non
  &\phantom{=\ }
  - m^2(1 - \sigma\cos\eta)
  - \epsilon m^2\bpi^3\sin\eta\ .
  \label{eq:Skyrme_Lagrangian_eta_rescaled}
\end{align}
We note that $\sigma$ is an auxiliary field due to the nonlinear
constraint $\sigma^2+\bpi\cdot\bpi=1$.
Regarding only $\bpi$ as the physical field, we have
\beq
\sigma = \sqrt{1-\bpi\cdot\bpi}
= 1 - \frac12\bpi\cdot\bpi + \mathcal{O}(\bpi^4)\ ,
\eeq
and therefore we can write down the quadratic Lagrangian as
\begin{align}
\Lag^{\rm quad} &=
-\frac12\p_\mu\eta\p^\mu\eta
-\frac12\p_\mu\bpi\cdot\p^\mu\bpi
-\frac12m^2(\bpi\cdot\bpi + \eta^2 + 2\epsilon \bpi^3\eta) \ .
\label{eq:static_lagrangian_asymp}
\end{align}
The addition of $\eta$ has produced an off-diagonal contribution to
the mass term, due to the mixing between $\eta$ and $\bpi^3$.
The new mass matrix reads:
\begin{equation}
M^2 = m^2 \begin{pmatrix}
1 & 0 & 0 & 0\\
0 & 1 & 0 & 0\\
0 & 0 & 1 & \epsilon\\
0 & 0 & \epsilon & 1
\end{pmatrix}.
\end{equation}
Its eigenvalues in units of $m^2$ are $1,1,1\pm\epsilon$ and the
eigenvectors corresponding to the two non-unity eigenvalues are given
by $\frac{1}{\sqrt{2}}(\bpi^3\pm \eta)$.
The quadratic Lagrangian \eqref{eq:static_lagrangian_asymp} can now be
rewritten in terms of 
$\widetilde\bpi=\big(\bpi^1,\bpi^2,\frac{1}{\sqrt{2}}(\bpi^3+\eta),\frac{1}{\sqrt{2}}(\bpi^3-\eta)\big)$ as:
\begin{equation}
\Lag^{\rm quad} =
-\frac{1}{2}\partial_\mu\widetilde\bpi\cdot\partial^\mu\widetilde\bpi
- \frac{1}{2}\widetilde\bpi\widetilde{M}^2\widetilde\bpi^{\rm T}\ ,
\label{eq:asymp_lagrangian_diag}
\end{equation}
where $\widetilde{M}^2$ is the diagonal form of $M^2$. 
\begin{equation}
\widetilde{M}^2 = m^2\diag\big(1,1, 1+\epsilon,1-\epsilon\big)\ .
\end{equation}

We want to solve linear equations in a spherical coordinate system  which reflects the behavior at large distance.
We end up with the following expressions:
\begin{equation}
\begin{split}
\bpi^1 = C\frac{mr + 1}{r^2}e^{-mr}\,\hat{x}^1\ ,\\ 
\bpi^2 = C\frac{mr + 1}{r^2}e^{-mr}\,\hat{x}^2\ ,
\label{eq:pi1_pi2}
\end{split}
\end{equation}
and we obtain for $\bpi^3$ and $\eta$:
\begin{equation}
\begin{split}
\bpi^3 &= \frac{D}{2}  \frac{m\sqrt{1+\epsilon}\, r + 1}{ r^2}	e^{-m\sqrt{1+\epsilon}\, r} \hat{x}^3+ \frac{E}{2}  \frac{m\sqrt{1-\epsilon}\, r + 1}{ r^2}	e^{-m\sqrt{1-\epsilon}\, r} \hat{x}^3\ ,\\
\eta &= \frac{D}{2}  \frac{m\sqrt{1+\epsilon}\, r + 1}{ r^2}	e^{-m\sqrt{1+\epsilon}\, r} \hat{x}^3 - \frac{E}{2}  \frac{m\sqrt{1-\epsilon}\, r + 1}{ r^2}	e^{-m\sqrt{1-\epsilon}\, r} \hat{x}^3\ ,
\label{eq:pi3_eta}
\end{split}
\end{equation}
where $C$, $D$ and $E$ are coefficients to be determined and we have used that
\beq
\p_i^2(f(r)\hat{x}^a) - m^2f(r)\hat{x}^a
= \left(f''(r) + \frac{2}{r}f'(r) - \frac{2}{r^2}f(r) - m^2f(r)\right)\hat{x}^a\ ,
\eeq
and hence $f(r)$ of Eq.~\eqref{lineartailspheric} is a solution to the
above equation for any $a=1,2,3$.
Now since we want to cover the 3-sphere target space, we point each of
the pions in the three respective Cartesian directions.
In order to let $\bpi^3$ be a small perturbation around the expected
solution, we choose the fourth tail, i.e.~that of $\widetilde{\bpi}^4$
to be pointed also in the $\hat{x}^3$-direction.
This way, $\bpi^3$ points purely in the $\hat{x}^3$ direction and is
only deformed radially by the presence of nonvanishing $\epsilon$. 

For a vanishing quark mass difference, i.e.~$\epsilon=0$, and/or
vanishing mass quark $m_q$, $C=D=E$ and this corresponds to the
asymptotic linear tail of the spherically symmetric Skyrmion.

The moment of inertia tensor is given by
\begin{align}
\I_{i j} = \int\d^3x\;\epsilon_{i l m}\epsilon_{j n p}x^lx^n\Big[&
  \p_m\eta\p_p\eta
  +\p_m\bpi\cdot\p_p\bpi\non
  &+\p_m\sigma\p_p\sigma+\big(\p_m\sigma\p_p\sigma+\p_m\bpi\cdot\p_p\bpi\big)
  \big(\p_k\sigma\p_k\sigma+\p_k\bpi\cdot\p_k\bpi\big)\non
  &-\big(\p_m\sigma\p_k\sigma+\p_m\bpi\cdot\p_k\bpi\big)
  \big(\p_p\sigma\p_k\sigma+\p_p\bpi\cdot\p_k\bpi\big)\Big]\ ,
\label{eq:Ical}
\end{align}
with the kinetic energy
\beq
E_{\rm rot} = \frac12 \Omega_i \I_{i j} \Omega_j\ ,
\eeq
where $\Omega_i$ is given in Eq.~\eqref{eq:angular_velocity_def} and
we have used the fact that under a time-dependent rotation
$\bx\to R(t)\bx$, we have
\beq
\p_0\eta(R\bx)
=\frac{\p\eta}{\p(R\bx)^k}\epsilon_{ijk}(R\bx)^j\Omega_i\ ,\qquad
\Omega_i = -\frac12\epsilon_{ijk}(\dot{R}R^{\rm T})_{jk}\ .
\eeq
Writing the coordinates as $R\bx\to\bx$ to avoid unnecessary clutter,
we arrive at the expression \eqref{eq:Ical}.

We now proceed with the computation of the moment of inertia of the
Skyrmion using the linear tail.
The strategy is the same we used in
Eqs.~\eqref{linearI0}-\eqref{linearI} for the spherical Skyrmion.
The linear tails of the fields are given in
Eqs.~\eqref{eq:pi1_pi2}-\eqref{eq:pi3_eta}.
As we are considering only the linear tail, we use only the part of
the inertia tensor that comes from the quadratic Lagrangian $\Lag_2$
(the first line of Eq.~\eqref{eq:Ical}). 
We compute first the contribution from $\bpi^3$ that has the form
$\bpi^3=\bpi^3(r,\theta)$ since it only depends on $r$ and
$\hat{x}^3$, obtaining: 
\beq
\I_{\pi^3} = \int\d r\d\theta\d\varphi\; r^2\sin\theta
\Big(\frac{\partial \bpi^3}{\partial \theta}\Big)^2  \begin{pmatrix}
\sin^2{\varphi}  & 0 & 0 \\
0 & \cos^2{\varphi}  & 0  \\
0 & 0 & 0  \\ 
\end{pmatrix}.
\eeq
Integrating over $\varphi$ and summing the two similar contributions from  
$\bpi^3$ and $\eta$ we have
\beq
\I_{\pi^3}+\I_{\eta} =  \pi\int\d r d\theta\; r^2 \sin\theta \left( \Big(\frac{\partial \bpi^3}{\partial \theta}\Big)^2 + \Big(\frac{\partial \eta}{\partial \theta}\Big)^2 \right)  \begin{pmatrix}
1 & 0 & 0 \\
0 & 1 & 0  \\
0 & 0 & 0  \\ 
\end{pmatrix}.
\eeq
Using the asymptotic linear solutions found in
Eqs.~\eqref{eq:pi1_pi2}-\eqref{eq:pi3_eta} we get
\begin{align}
\I_{\pi^3}+\I_{\eta} = \frac{8\pi}{3}\int\d r\; r^2 \bigg(
& \Big(\frac{D}{2}  \frac{m\sqrt{1+\epsilon}\, r + 1}{ r^2}
e^{-m\sqrt{1+\epsilon}\, r}  \Big)^2\non
&+\Big(\frac{E}{2}  \frac{m\sqrt{1-\epsilon}\, r + 1}{ r^2}
e^{-m\sqrt{1-\epsilon}\, r}  \Big)^2 \bigg)
\begin{pmatrix}
1 & 0 & 0 \\
0 & 1 & 0  \\
0 & 0 & 0  \\ 
\end{pmatrix}.
\label{intlin1} 
\end{align}
Using the asymptotic linear solutions found in
Eqs.~\eqref{eq:pi1_pi2}-\eqref{eq:pi3_eta}, the contributions from
$\bpi^1$ and $\bpi^2$ are:
\beq
\I_{\pi^1} + \I_{\pi^2} =  \frac{4\pi}{3}\int\d r\; r^2 \Big( C
\frac{m r + 1}{ r^2}	e^{-m  r}   \Big)^2
\begin{pmatrix}
1 & 0 & 0 \\
0 & 1 & 0  \\
0 & 0 & 2  \\ 
\end{pmatrix}.
\label{intlin2}
\eeq
Summing over all contributions, we finally obtain the total inertia
tensor 
\beq
\I = \I_{\pi^1} + \I_{\pi^2} + \I_{\pi^3} + \I_{\eta}\ .
\eeq
The inertia tensor can be split into an identity part and a
deformation 
\beq
\I = I\,\mathbb{1}_3 + \delta \begin{pmatrix}
1 & 0 & 0 \\
0 & 1 & 0  \\
0 & 0 & -2  \\ 
\end{pmatrix}.
\label{intenssplit}
\eeq

The moment of inertia of the resulting Skyrmion may be more oblate
like a pancake or more prolate like an American football,
depending on the sign of $\delta$. 
For $\epsilon=0$ and $C=D=E$, we obtain $\delta=0$ and
\beq
I =  \frac{8\pi}{3}\int\d r\; r^2 C^2\left(\frac{m r + 1}{ r^2} e^{-m r}\right)^2,
\eeq
which is the result of Eq.~\eqref{linearI0} for the spherical case.

These integrals, as in the spherical case, are UV divergent. 
As before, these solutions are valid up to a UV cutoff $R_{\rm UV}$
where the solution becomes nonlinear; $R_{\rm UV}$ is essentially the
size of the Skyrmion $R_{\rm Sk}$.
Integrating over $r$ from the UV cutoff, we obtain
\begin{align}
  \I &= \frac{4\pi}{3}C(m)^2e^{-2 m R_{\rm UV}}
  \left(\frac{m}{2}+\frac{1}{R_{\rm UV}}\right)
  \begin{pmatrix}
    1 & 0 & 0 \\
    0 & 1 & 0  \\
    0 & 0 & 2  \\ 
  \end{pmatrix}\non
  &\phantom{=\ }
  +\frac{2\pi}{3}\bigg[
    D(m)^2e^{-2 m\sqrt{1+\epsilon} R_{\rm UV}}
    \left(\frac{m\sqrt{1+\epsilon}}{2} + \frac{1}{R_{\rm UV}}\right)\non
  &\phantom{=+\frac{2\pi}{3}\bigg[\ }
    +E(m)^2e^{-2 m\sqrt{1-\epsilon} R_{\rm UV}}
    \left(\frac{m\sqrt{1-\epsilon}}{2} + \frac{1}{R_{\rm UV}}\right)
    \bigg]
  \begin{pmatrix}
    1 & 0 & 0 \\
    0 & 1 & 0  \\
    0 & 0 & 0  \\ 
  \end{pmatrix}.
\end{align}
Expanding in $m$ yields a UV divergent term $\propto1/R_{\rm UV}\sim1/R_{\rm Sk}$:
\begin{align}
  \I &= \frac{4\pi}{3}\frac{C(0)^2}{R_{\rm UV}}
  \begin{pmatrix}
    1 & 0 & 0 \\
    0 & 1 & 0  \\
    0 & 0 & 2  \\ 
  \end{pmatrix}
  +\frac{2\pi}{3}\frac{\left(D(0)^2+E(0)^2\right)}{R_{\rm UV}}
  \begin{pmatrix}
    1 & 0 & 0 \\
    0 & 1 & 0  \\
    0 & 0 & 0  \\ 
  \end{pmatrix}\non
  &\phantom{=\ }
  -2\pi C(0)^2 m
  \begin{pmatrix}
    1 & 0 & 0 \\
    0 & 1 & 0  \\
    0 & 0 & 2  \\ 
  \end{pmatrix}
  -\pi\left(\sqrt{1+\epsilon}D(0)^2 + \sqrt{1-\epsilon}E(0)^2\right)m
  \begin{pmatrix}
    1 & 0 & 0 \\
    0 & 1 & 0  \\
    0 & 0 & 0  \\ 
  \end{pmatrix}\non
  &\phantom{=\ }
  +\mathcal{O}(m^2,m^2\epsilon)\ ,
\end{align}
where we have assumed that $C'(0)=D'(0)=E'(0)=0$.
Composing the tensor into the diagonal and the eighth Gell-Mann
generator according to Eq.~\eqref{intenssplit}, we get
\begin{align}
  I &= \frac{4\pi}{9}\frac{D^2+E^2+4C^2}{R_{\rm UV}}
  -\frac{2\pi}{3}\left(\sqrt{1+\epsilon}D^2 + \sqrt{1-\epsilon}E^2 + 4C^2\right)m
  +\mathcal{O}(m^2,m^2\epsilon)\ ,\label{eq:Imexp}\\
  \delta &= \frac{2\pi}{9}\frac{D^2+E^2-2C^2}{R_{\rm UV}}
  -\frac{\pi}{3}\left(\sqrt{1+\epsilon}D^2 + \sqrt{1-\epsilon}E^2 - 2C^2\right)m
  +\mathcal{O}(m^2,m^2\epsilon)\ ,
  \label{eq:deltamexp}
\end{align}
where in the above expressions $C=C(0)$, $D=D(0)$, and $E=E(0)$ and
$R_{\rm UV}$ can be approximated by $R_{\rm Sk}$.

If we set $D(0)=E(0)=C(0)$, the tensor simplifies as
\begin{align}
  I &= \frac{8\pi}{3}\frac{C(0)^2}{R_{\rm UV}}
  -\frac{2\pi}{3} C(0)^2\left(4 + \sqrt{1+\epsilon} + \sqrt{1-\epsilon}\right)m
  +\mathcal{O}(m^2,m^2\epsilon)\ ,\label{eq:Imexp_CDEequal}\\
  \delta &= \frac{\pi}{3}C(0)^2\left(2 - \sqrt{1+\epsilon} - \sqrt{1-\epsilon}\right)m
  +\mathcal{O}(m^2,m^2\epsilon,m\epsilon^4)\ ,
  \label{eq:deltamexp_CDEequal}
\end{align}
which is difficult to establish without knowledge of the full
nonlinear solutions.
Expanding the square roots in the linear term of $I$, we obtain
\begin{align}
I(m) &= I(0)
  -4\pi C(0)^2\left(1 - \frac{\epsilon^2}{24}\right)m
  +\mathcal{O}(m^2,m^2\epsilon,m\epsilon^4)\ ,
  \label{eq:Imexp_CDEequal_epsilonexp}\\
\delta &= \frac{\pi}{12}C(0)^2\epsilon^2 m
  +\mathcal{O}(m^2,m^2\epsilon,m\epsilon^4)\ .
  \label{eq:deltamexp_CDEequal_epsilonexp}
\end{align}
One could make the assumption that
$|\bpi^1|_{r=R_{\rm UV}}\simeq|\bpi^2|_{r=R_{\rm UV}}\simeq|\bpi^3|_{r=R_{\rm UV}}$
and $|\eta|_{r=R_{\rm UV}}\ll|\bpi^3|_{r=R_{\rm UV}}$;
the latter assumption is equivalent with assuming $D\sim E$.
Equating the magnitudes of the tails
\begin{align}
C(m)\frac{m R_{\rm Sk} + 1}{R_{\rm Sk}^2}e^{-mR_{\rm Sk}} &\simeq
D(m)\frac{m\sqrt{1+\epsilon}\,R_{\rm Sk} + 1}{R_{\rm Sk}^2} e^{-m\sqrt{1+\epsilon}\, R_{\rm Sk}}\non&\simeq
E(m)\frac{m\sqrt{1-\epsilon}\, R_{\rm Sk}+ 1}{R_{\rm Sk}^2} e^{-m\sqrt{1-\epsilon}\, R_{\rm Sk}}\ ,
\end{align}
which when expanding to second order in $m$ yields
\beq
D-C\simeq \frac12m^2\epsilon R_{\rm Sk}^2 C\ , \qquad
E-C\simeq -\frac12m^2\epsilon R_{\rm Sk}^2 C\ .
\label{eq:CDEsplitting}
\eeq
Plugging this approximation into
Eqs.~\eqref{eq:Imexp}-\eqref{eq:deltamexp}, we obtain exactly the
results
\eqref{eq:Imexp_CDEequal_epsilonexp}-\eqref{eq:deltamexp_CDEequal_epsilonexp},
with the order $m^0$ terms generating $m^4\epsilon$ corrections and
the order $m$ terms generating $m^3\epsilon$ corrections, all of which
we neglect to leading order.
This shows that the results
\eqref{eq:Imexp_CDEequal_epsilonexp}-\eqref{eq:deltamexp_CDEequal_epsilonexp}
are quite robust.

Note that the order $m$ correction to $I$ is insensitive to the cutoff
$R_{\rm Sk}$ unlike the leading-order term; we can therefore trust the
coefficient in front this linear-in-$m$ term.
Notice also that we need to take $\epsilon^2$ into account, but we
have expanded only to the first order in $m$.
This is consistent in the limit
\beq
m \ll |\epsilon|\ .
\label{mepcond}
\eeq
In reality, the physical parameters are $m\simeq 0.3$ and
$\epsilon=-0.34$, so for $I$ one should also take the higher-order
correction terms of order $m^2$ and $m^2\epsilon$ into account, which
however are more complicated expressions and will depend also on the
double derivatives $C''(0)$, $D''(0)$ and $E''(0)$.

The form of the inertia tensor is that of Eq.~\eqref{intenssplit}.
This is an axially symmetric rotor, it may be oblate or prolate
according to whether $\delta$ is negative or positive.
The related quantum Hamiltonian becomes then:
\begin{equation}
\begin{split}
  H_{\rm rot} &= \frac{1}{2}\I^{-1}_{ij}J_i J_j
  = \frac{1}{2}\frac{J^2}{I + \delta} + \frac{1}{2}J^2_3\left[\frac{1}{I -2 \delta} - \frac{1}{I + \delta}\right]\\ 
& = \frac{1}{2}\frac{1}{I + \delta}j(j+1) + \frac{3}{2}\frac{\delta}{(I+\delta)(I - 2 \delta)} j^2_3 \ ,
\end{split}
\label{eq:H_rot}
\end{equation}
where $j(j+1)$ and $j_3$ are the eigenvalues of the $J^2$ and $J_3$
operators, respectively.
Eq.~\eqref{eq:H_rot} distinguishes between different values of
$j_3\in[-j,\ldots,j]$ but not their sign, so the states that
correspond to the fundamental representation with $j=1/2$ are left
degenerate, i.e.~the proton and the neutron have the same mass.
The $j=3/2$ ones instead get split into two pairs (i.e.~$\Delta^{++}$
and $\Delta^-$ with $|j_3|=3/2$ whereas $\Delta^{+}$ and $\Delta^0$ with
$|j_3|=1/2$):
\beq
M_{\Delta^{++},\Delta^-} - M_{\Delta^{+} ,\Delta^0}
=\frac{1}{2} (M_{\Delta^{++}} - M_{\Delta^{+}}  -M_{\Delta^{0}} + M_{\Delta^{-}})\ ,
\eeq
where for the spin Hamiltonian \eqref{eq:H_rot} we have
\begin{align}
  M_{\Delta^{++},\Delta^-}&=M_{\Delta^{++}}=M_{\Delta^-}=\left.M_{\rm Sk}^{\rm static}+H_{\rm rot}\right|_{j=\tfrac32,|j_3|=\tfrac32}\ ,\\
  M_{\Delta^{+},\Delta^0}&=M_{\Delta^{+}}=M_{\Delta^0}=\left.M_{\rm Sk}^{\rm static}+H_{\rm rot}\right|_{j=\tfrac32,|j_3|=\tfrac12}\ ,
  \end{align}
since $H_{\rm rot}$ of Eq.~\eqref{eq:H_rot} does not distinguish
between $j_3$ positive and negative, whereas $M_{\rm Sk}^{\rm static}$ is the static soliton mass.
In particular
\beq
\frac{1}{2} (M_{\Delta^{++}} - M_{\Delta^{+}}  -M_{\Delta^{0}} + M_{\Delta^{-}})
=\frac{3\delta}{(I+\delta)(I - 2 \delta)}\simeq \frac{3\delta }{I^2}\ .
\eeq
Assuming the previous results are valid we get
\beq
\frac{1}{2} (M_{\Delta^{++}} - M_{\Delta^{+}}  -M_{\Delta^{0}} + M_{\Delta^{-}})
\simeq \frac{\pi C(0)^2m}{I^2}
\left(2 - \sqrt{1+\epsilon} - \sqrt{1-\epsilon}\right)\ .
\eeq
For the numerical values $m\simeq 0.3$ and $\epsilon\simeq-0.34$ we
obtain
\beq
\frac{1}{2} (M_{\Delta^{++}} - M_{\Delta^{+}}  -M_{\Delta^{0}} + M_{\Delta^{-}})
\simeq  f_{\pi} e^3 \,  4.6 \times 10^{-5} \simeq 0.54\MeV \ .
\eeq
Note that in the small-$m$ limit, the linear deviation
\eqref{eq:deltamexp_CDEequal} (positive $\delta$) makes the rotor
prolate, thus $\Delta^{++}$ and $\Delta^-$ have a higher mass.
At a finite but not small $m$, this may change and only a numerical
computation can tell if it is oblate or prolate; in fact, see below.
We saw that the small-$m$ condition is very good for the quantity
$I$.
For $\delta$ we have to be more careful, because also the condition
\eqref{mepcond} has to be respected.
So the core contribution to the splitting may be important for the
phenomenological values of $m$ and $\epsilon$.

We will now turn to the numerical computation of the Skyrme model with
the $\eta$ taken into account.
We use the numerical method developed for the $\omega$-Skyrme model in
Ref.~\cite{Gudnason:2020arj}, where we solve the equation of motion
for the scalar $\eta$ using the conjugate gradients method at every
step and solve the nonlinear equation of motion for the pions and the
sigma using arrested Newton flow.
The simplicity of the $\eta$ equation and the fact that $\eta$ is not
normalized, unlike the fact that $\sigma^2+\bpi^2=1$ (coming from
$\det U=1$), makes the conjugate gradients method very efficient.
The arrested Newton flow is implemented with a second order real-time
evolution of the field equations (ignoring the time dependence in the
Skyrme term), monitoring the (static) potential energy at every step
and setting all time derivatives to zero when the potential energy
increases with respect to the previous step.
This method implies that the fields are accelerating down towards the
minimal energy solution, but are ``gradient flowing'' up to the
minimum.
The specific implementation of the method was done in CUDA C for
NVIDIA GPUs and the lattice was chosen as a $160^3$ cubic lattice with
step size $0.088$ and a 5-point fourth-order finite difference stencil
for the derivatives. 

\begin{figure}[!ht]
  \centering
  \includegraphics[width=0.7\linewidth]{{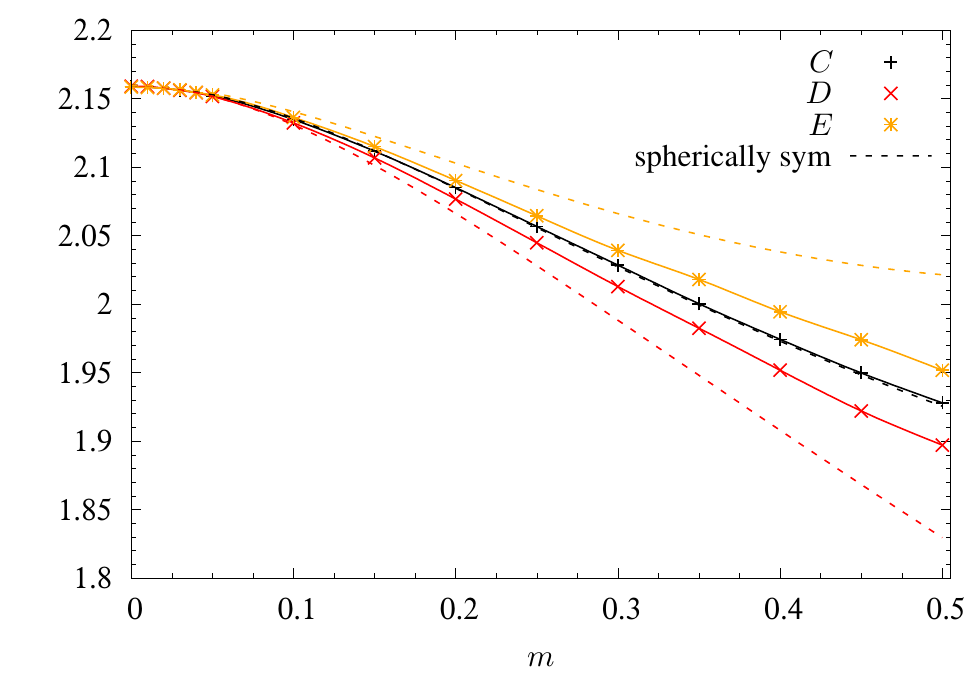}}
  \caption{The coefficients \eqref{eq:pi1_pi2}-\eqref{eq:pi3_eta} of
    the exponential tails of the pion and $\eta$ fields as functions
    of $m$. The black dashed curve is the result of the spherically
    symmetric Skyrmion (for which there is only one coefficient $C$ of
    all the three pions, whereas the yellow (upper) and red (lower)
    dashed curves are the prediction \eqref{eq:CDEsplitting}, which is
    correct in sign, but not quite in magnitude.
    In this figure $\epsilon=-0.34$, i.e.~its physical value.
  }
  \label{fig:CDE}
\end{figure}
The small-$m$ limit makes the numerics very hard compared to the usual
physical case of a finite pion mass parameter, often taken to be of
order one, ensuring a fast exponential decay and allowing numerical
computations with finite differences to be put on a finite box without
loss of precision.
Instead, we consider the numerics for $m$ in the range between typical
physical values and all the way down to zero, calling for a
modification of the usual methods.
We implement Dirichlet boundary conditions on a sufficiently large box,
dictated by the linearized exact solutions
\eqref{eq:pi1_pi2}-\eqref{eq:pi3_eta} and using as an initial guess,
the values $C=D=E$ with $C$ from the spherically symmetric case of the
previous section.
By trial-and-error we observe that only after a long simulation time,
the derivatives converge to their expected values dictated by
Eqs.~\eqref{eq:pi1_pi2}-\eqref{eq:pi3_eta} and hence we simply read
off the coefficients $C$, $D$ and $E$ from the derivatives at the end
of the simulation and start it again with the updated values of the
coefficients of the tails of the fields.
Iterating about 5 times gives a reasonably good precision on the
coefficients and the result is shown in Fig.~\ref{fig:CDE}.
First of all we observe that the behavior of the coefficients is
quadratic in $m$ for small $m$, as expected from the spherically
symmetric case.
Second of all, we confirm that $C(0)=D(0)=E(0)$ as it should be (since
$m=0$ turns off any splitting, viz.~there are no factors of $\epsilon$
without at least one power of $m$).
Third of all, we can confirm that the predicted splitting of the
coefficients follow Eq.~\eqref{eq:CDEsplitting}, i.e.~that $E>C>D$
for $m>0$ (recall that $\epsilon<0$).
The magnitude of the splitting is however not accurate, see the yellow
and red dashed lines in Fig.~\ref{fig:CDE}.

\begin{figure}[!htp]
  \centering
  \mbox{\subfloat[]{\includegraphics[width=0.49\linewidth]{{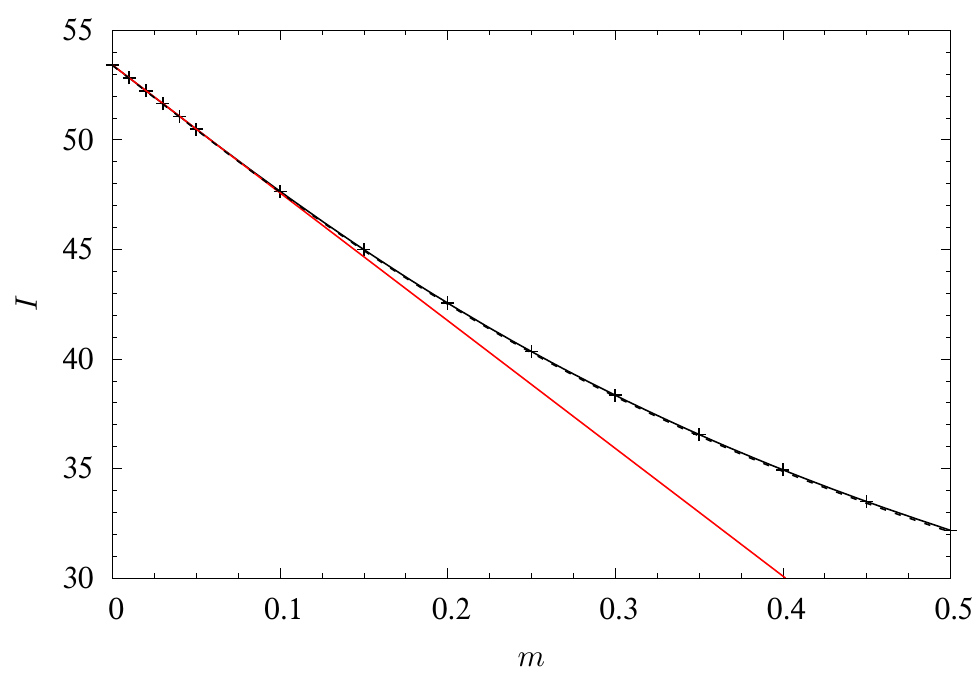}}}
    \subfloat[]{\includegraphics[width=0.49\linewidth]{{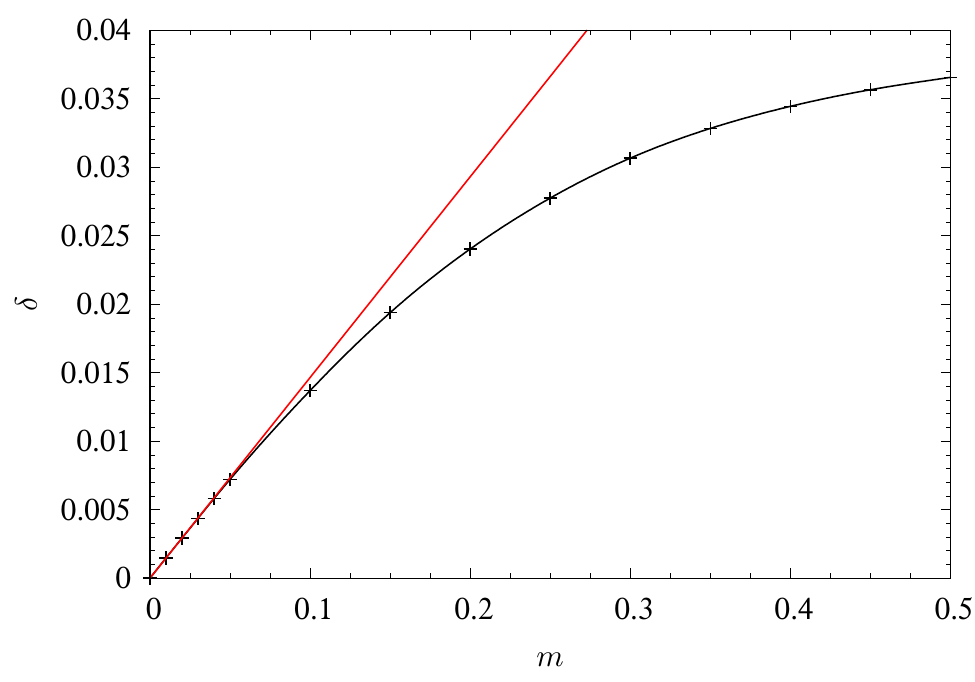}}}}
  \caption{(a) The trace part and (b) the traceless part (proportional
    to the Gell-Mann $\lambda^8$ generator, see
    Eq.~\eqref{intenssplit}) of the moment of inertia tensor as a
    function of $m$ computed by a full 3-dimensional computation on
    the lattice with the contribution from outside the lattice to
    infinity computed from the linearized tail solutions.
    The red straight lines are the predictions of
    Eq.~\eqref{eq:Imexp_CDEequal} and \eqref{eq:deltamexp_CDEequal},
    respectively, for panel (a) and (b). 
  }
  \label{fig:moi_delta}
\end{figure}
Since we cannot perform the numerical computations of sufficiently
large lattices in order to capture the accurate moment of inertia
tensor in the limit of small-$m$, we compute the moment of inertia
tensor on the largest sphere fitting into the lattice and calculate
the contribution from the outside by using the tails
\eqref{eq:pi1_pi2}-\eqref{eq:pi3_eta} and the coefficients of
Fig.~\ref{fig:CDE} -- the latter part is semi-analytic, although we
have to compute the $r$-$\theta$ integral numerically too.

We confirm that the linear prediction of both the trace part as well
as the splitting part ($\delta$) of the moment of inertia tensor is in
accord with Eqs.~\eqref{eq:Imexp_CDEequal} and
\eqref{eq:deltamexp_CDEequal}, at least up to numerical accuracy and
order $m^2$ corrections, see Fig.~\ref{fig:moi_delta}.
From the result that $\delta$ is everywhere positive, we can conclude
for the range of $m$ studied here, that the nucleon is always prolate.

\begin{figure}[!htp]
  \centering
  \includegraphics[width=\linewidth]{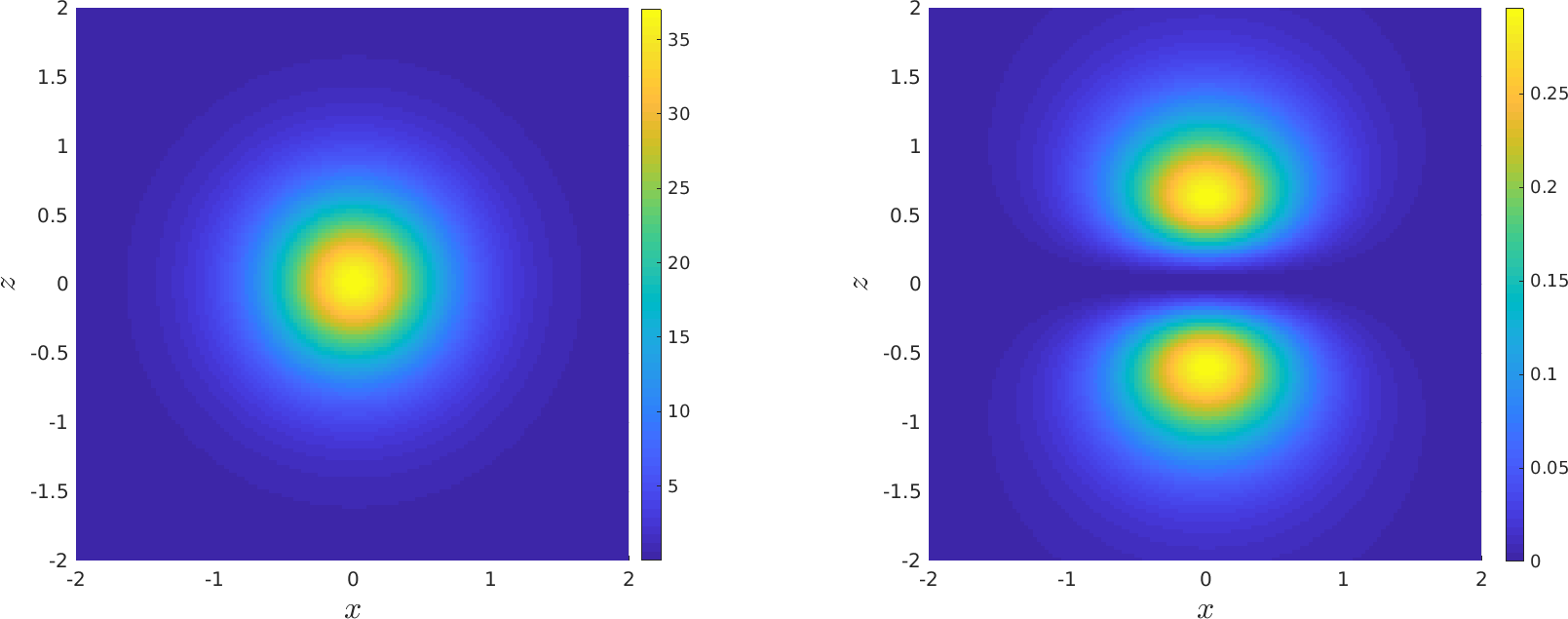}
  \caption{Left panel: The energy density $\mathcal{E}^{\rm tot}(x,0,z)$ of the nucleon, i.e.~the energy with spin contribution corresponding to $j=j_3=1/2$.
    Right panel: $\mathcal{E}^{\rm tot}(x,0,z)-\mathcal{E}(x,y,0)$.
    In this figure $m=0.3$ and $\epsilon=-0.34$.
  }
  \label{fig:endens}
\end{figure}
We plot the total energy density of the numerical computation, with
the spin contribution to the energy corresponding to the nucleon, in
Fig.~\ref{fig:endens},
where the total energy density in dimensionless units, is defined as
\beq
\mathcal{E}^{\rm tot}
= -\frac{e}{f_\pi}\Lag
+ \frac{e^4}{2}\left(\frac{j(j+1)\calI_{11}}{I_{11}^2} + \frac{j_3^2\calI_{33}}{I_{33}^2} - \frac{j_3^2\calI_{11}}{I_{11}^2}\right)\ ,\label{eq:Edensity_tot}
\eeq
with $-\Lag$ of Eq.~\eqref{eq:Skyrme_Lagrangian_eta_rescaled},
$\calI_{ij}$ of Eq.~\eqref{eq:Ical}, $I_{ij}=\int\d^3x\,\calI_{ij}$
and for the nucleon $j=|j_3|=\frac12$.
This energy density is constructed such that its integral gives the
energy of the quantized Hamiltonian \eqref{eq:H_rot} plus the
classical Skyrmion mass. 
From this figure, the nucleon appears to be prolate, but this is the
shape of the energy density near the core of the Skyrmion/nucleon.
We find the same shape for the Deltas.
\begin{figure}[!htp]
  \centering
  \includegraphics[width=\linewidth]{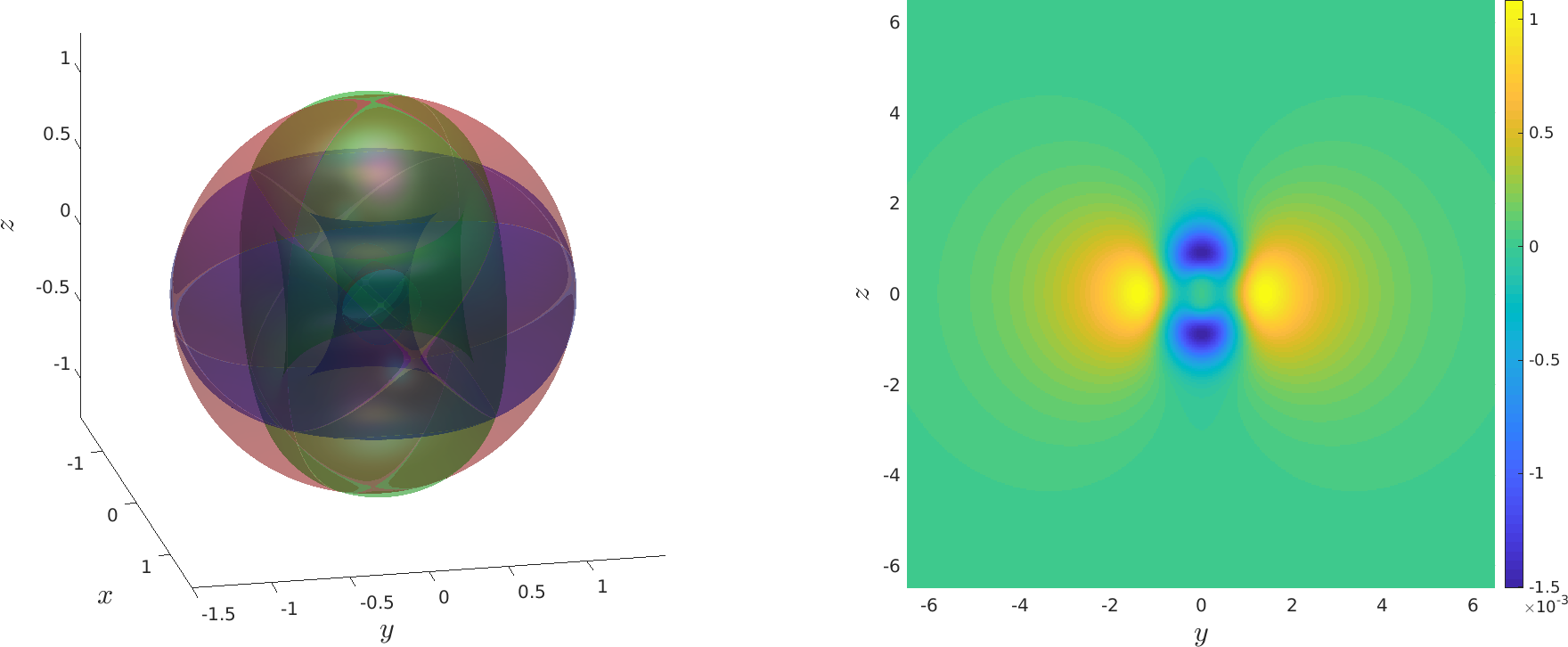}
  \caption{Left panel: The 3 diagonal components of the moment of
    inertia tensor density $\calI_{ii}$ with $i=1,2,3$ (not summed over) shown
    in red, green and blue respectively.
    Right panel: $\calI_{11}(x,0,z)-\calI_{33}(x,y,0)$.
    In this figure $m=0.3$ and $\epsilon=-0.34$.
  }
  \label{fig:inertia_diff}
\end{figure}
In Fig.~\ref{fig:inertia_diff}, we plot the 3 diagonal components of
the inertia tensor density of the Skyrmion, which are all tori about their
corresponding axis; e.g.~the $\calI_{11}$ component takes the shape of a torus
with the main axis in the $x^1$ direction and so on.
In the right panel of the figure, we show the $(y,z)$-plane of the
$\calI_{11}$ component with the $(x,y)$-plane of the $\calI_{33}$ component of
the inertia tensor density subtracted off.
This latter panel illustrates that far from the core, the tail effect
of the $\calI_{11}$ (and $\calI_{22}$ is similar) gives a larger
positive relative contribution (in yellow) than the $\calI_{33}$ gives
a negative relative contribution (in blue).
The combined effect is that the tail of the Skyrmion makes it prolate.
This effect dominates the oblate property of the core.

The presence of the new $\eta$ particle has affected the Skyrmion
quite a bit, but we still cannot obtain the mass splitting between the
neutron and the proton.  

By making the spin-isospin association, as in the \textit{hedgehog}
example, we can infer that, the residual neutron-proton symmetry we
just found, is strictly related to how the Skyrmion rotates.
Eq.~\eqref{eq:H_rot} reveals that any left or right rotating Skyrmion
has the same energy, which instantly translates into the proton and
the neutron being the same soliton solution rotating in opposite
directions.

\begin{figure}[!ht]
  \centering
  \includegraphics[width=0.7\textwidth]{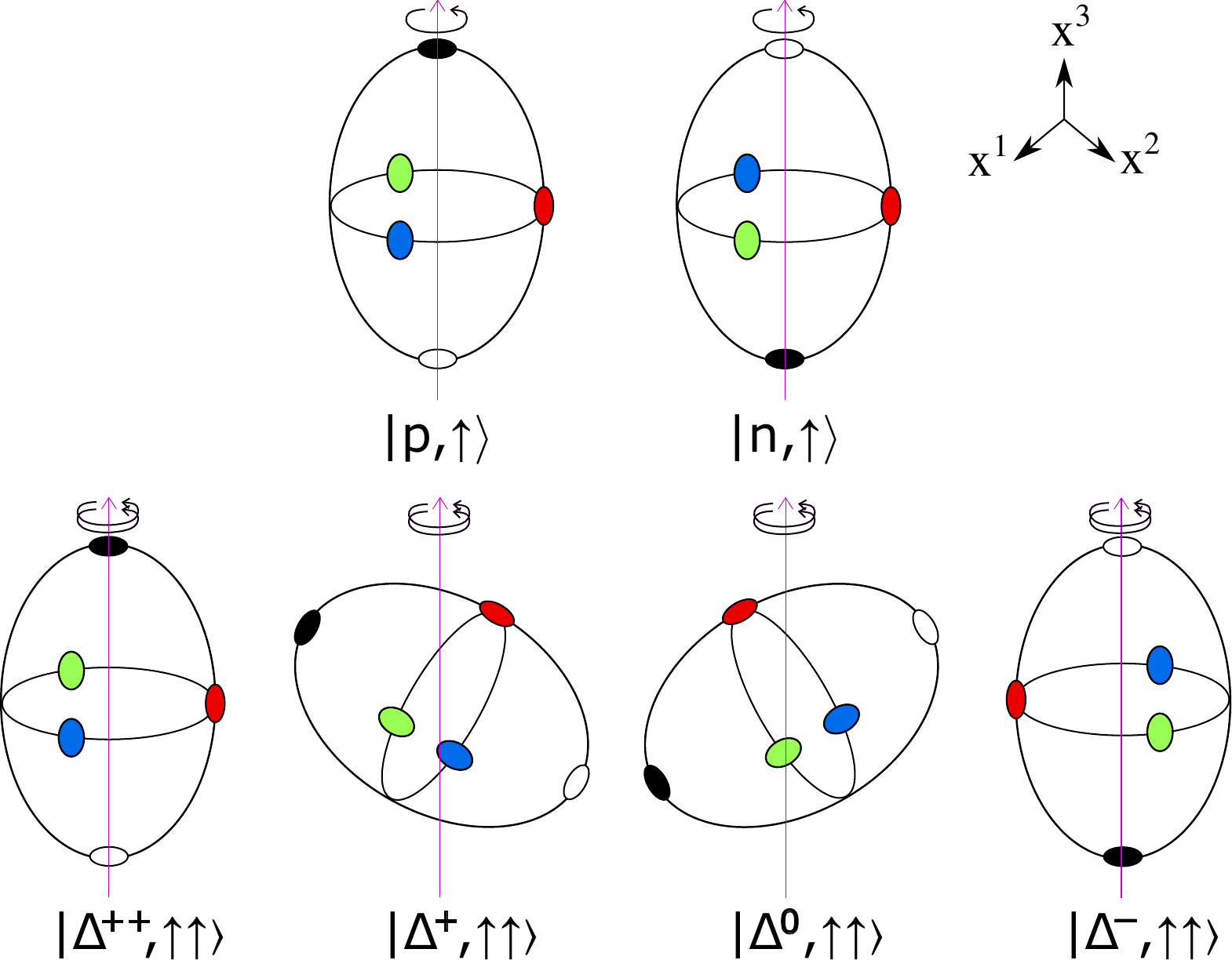}
  \caption{Schematic representations of the different states corresponding to the
    fundamental and $\Delta$ representation of $\SU(2)$,
    all with maximal spin-up, with the modification of shape due to
    cylindrical symmetry of the solutions.
    The colors are the same as in Fig.~\ref{fig:4_states}.
  }
  \label{fig:4_states_football}
\end{figure}

The shape of the resulting Skyrmion is obviously dependent upon who
wins, but besides that nothing changes the fact that the neutron and
the proton do not get different masses, as they are the same state
rotating in different $\SU(2)_I$ directions, as illustrated in
Fig.~\ref{fig:4_states_football}. 
We shall later come back to these terms along with all the others,
which give an extra contribution to the moment of inertia, but for now
we should finally address the latter approach, where the splitting
is coming from vector mesons.

The persistence of the degeneracy of the neutron-proton multiplet, or
in general of all the states with the same $|I_3|$, can also be
understood by a more general symmetry argument.
The Lagrangian with the pions, $\eta$ and the mass splitting, breaks the
continuous isospin but leaves invariant a discrete isospin parity, for
example as the reflection $R_{\rm iso, 13}: U \to U^{\rm T}$ which
acts as $(\pi^1,\pi^2,\pi^3) \to (\pi^1,-\pi^2,\pi^3)$.
To pass from the proton to the neutron with both spin up (see the
first line of Fig.~\ref{fig:4_states_football}) we can use two
symmetry transformations, a space parity
$P_{\rm spc} : (x^1,x^2,x^3)\to  (-x^1,-x^2,-x^3)$ combined with the
isospin parity $R_{\rm iso, 13}$.
These discrete symmetries protect the degeneracy.
This symmetry argument works independent of the various
approximations made above, for example the rigid rotor and the
semiclassical quantization.

\section{Isospin splitting in the Witten-Sakai-Sugimoto model}
\label{tre}

Isospin multiplets turn out to be completely degenerate if we include
only pions in such effective theory with $N_f=2$, while it is possible
to have a partial mass splitting between the Deltas if we add the
$\eta$ meson.
To completely remove the degeneracy, obtaining also a splitting in
mass between the nucleon states, i.e.~the proton and the neutron, the
inclusion of vector mesons is decisive.
The Sakai-Sugimoto model \cite{Sakai:2004cn,Sakai:2005yt} has the
advantage of automatically including all vector fields and their
excitations in a consistent way. 
Thus it can account for the presence of the neutron-proton mass
splitting incorporating the effect previously considered in
Refs.~\cite{Ebrahim:1987mu,Epele:1988ak,Jain:1989kn,Speight:2018zgc}.
Our aim is to incorporate both effects on the Deltas: The one studied in
Ref.~\cite{Bigazzi:2018cpg} that also produced the proton-neutron mass
splitting, and the one developed in Sec.~\ref{due} due to the
splitting of the moments of inertia, and see the complete splitting of
the spectrum of Deltas.

We briefly present the  baryons in the holographic model of
Sakai-Sugimoto.
We focus on the low-energy flavor dynamics on the fixed metric
background in 5-dimensions, as presented in
Refs.~\cite{Hata:2007mb,Hashimoto:2008zw}.
The gauge fields on the D8-branes are expanded on the $\U(2)$ basis as
\begin{equation}\label{notationsu2}
\mathcal{A} = \hA \frac{\mathbb{1}_2}{2}+A^a\frac{\tau^a}{2}\ ,  \qquad 
\mathcal{F} = \hF\frac{\mathbb{1}_2}{2} + F^a \frac{\tau^a}{2}\ .
\end{equation}
The 5D reduced metric and the effective action are given by
\begin{align}
\d s^2 &=  \frac{1}{h(z)^2} \d x^{\mu} \d x_{\mu} +  h(z)^2 \d z^2\ ,\\
\label{SYM}
S_{D8} &= - \kappa\Tr\int\d^4x\d z \left[k(z)\mathcal{F}_{\mu z}\mathcal{F}^{\mu z} + \frac{1}{2}h(z)\mathcal{F}_{\mu\nu}\mathcal{F}^{\mu\nu}\right] + S_{\rm CS}\ ,
\end{align}
with $h(z),k(z)$ being two functions who play the role of warp factors
amounting to
\begin{equation}\label{defhk}
  k(z)   = 1+z^2 \ , \qquad
  h(z)   = \frac{1}{(1+z^2)^{1/3}}\ .
\end{equation}
We can see that on the D8-brane's world volume lives a
five-dimensional Yang-Mills gauge theory with nontrivial warp
factors.
It will sometimes be useful to see the explicit $N_c$ dependence of
the overall constant of the action, hence we define
\beq
\kappa =a  \lambda N_c \ , \qquad
a= \frac{1}{ 216\pi^3}\ .
\eeq 
The Chern-Simons action has the explicit expression:
\begin{equation}\label{SCSDefintive}
S  _{\rm CS}=  \frac{N_c}{384 \pi^2} \epsilon^{\alpha_1 \alpha_2 \alpha_3 \alpha_4 \alpha_5} \int\d^4x\d z\;\hA_{\alpha_1} \left[ 3F_{\alpha_2 \alpha_3}^a F_{\alpha_4 \alpha_5}^a +\hF_{\alpha_2 \alpha_3}\hF_{\alpha_4 \alpha_5}\right].
\end{equation}
Everything is in the dimensional units of the Kaluza-Klein
compactification scale $M_{\rm KK}$, so the model has two
dimensionless couplings: $N_c$ and $\lambda$ (or $\kappa$) and a mass
scale $M_{\rm KK}$.

The holonomy of the $\mathcal{A}_z$ field can be interpreted as the
pseudoscalar matrix in a chiral theory \cite{Sakai:2004cn}, so that we
will adopt the following definitions
\beq\label{AzU}
\mathcal{P}\exp\left(\frac{\i}{2}\int\d z \hA_z\right) \equiv e^{\i\varphi(x)}\ , \qquad
\mathcal{P}\exp\left(\i\int\d z A_z\right) \equiv U(x)\ .
\eeq
Finally, the model can be supplemented with quark masses via the
inclusion of the Aharony-Kutasov action \cite{Aharony:2008an}:
\beq \label{SAK}
S_{\rm AK} = c\int\d^4x \Tr \mathcal{P}\left[Me^{\i\int\d z \mathcal{A}_z} -\mathbb{1}_2  + {\rm h.c.}\right].
\eeq
Note that action \eqref{SAK} exhibits nonlocality in the holographic
direction. 
$M $ is the quark mass matrix and $c$ is a constant (related to the
chiral condensate) that we can fix to reproduce the expected chiral
perturbation theory Lagrangian: This parameter is determined using the
pion mass as an input by the Gell-Mann-Oakes-Renner relation
\begin{equation}\label{cvalue}
c(m_u+m_d) = \frac{1}{2}f_{\pi}^2 m^2\ ,
\end{equation}
with $f_{\pi}$ being the pion decay constant and $m$ the pion mass in
units of $M_{\rm KK}$.
Employing definitions \eqref{AzU} we can cast the quark mass action
into the form
\beq\label{SAKU}
S_{\rm AK} = c\int\d^4x\Tr\left[Me^{\i\varphi}U -\mathbb{1}_2 + {\rm h.c.}\right].
\eeq
This is exactly the same as we had for the Skyrme model.
In fact, from the point of view of the quark mass and mass splittings,
the Skyrme model and the WSS model are exactly the same, with the
caveat that in the WSS model vector mesons and the $\eta$ are
automatically included, and the couplings are fixed by the two free
parameters $\lambda$, $M_{\rm KK}$.
In the Skyrme truncation of the model, we have \cite{Sakai:2004cn}
\beq
\frac{\kappa}{\pi} = \frac{f^2_{\pi}}{4}\ , \qquad
\frac{\kappa c_S}{2} = \frac{1}{32e^2}\ ,\qquad c_S \simeq 0.156 \ .
\eeq
One popular way to calibrate the parameters with mesons is
\beq
M_{\rm KK} = 949\MeV\ , \qquad
\kappa = 0.00745\ , \qquad
N_c=3\ , \qquad
\lambda = 16.63\ .
\label{calibration}
\eeq
This calibration is done by fitting the rho-meson mass $M_{\rho} =
776\MeV$ and the pion decay constant $f_{\pi}=92.4\MeV$. This mesonic
fit is, however, not the only possibility, and in particular it
overestimates masses in the baryon sector.

In the context of this holographic model, a baryon is realized as an
instantonic configuration of the gauge field $\mathcal{A}$.
The BPST instanton is given by
\begin{equation}
\label{BPST}
A_M = -\i f(\xi) g\partial_Mg^{-1},
\end{equation}
where we have introduced the following functions:
\begin{equation}
f(\xi) = \frac{\xi^2}{\xi^2 + \rho^2}\ , \qquad
g(\bx,z)= \frac{\left(z-Z\right)-\i(\bx-\bX)\cdot\btau}{\xi}\ ,
\end{equation}
the radial coordinate in 3-space including the holographic direction,
i.e.~the $(\bx,z)$-space:
\beq
\xi = \sqrt{(\bx - \bX)^2 + (z-Z)^2}\ ,
\eeq
and the parameters $(\bX,Z,\rho)$ describing the center and the size of the instanton in the $(\bx,z)$-space. 
The field strengths associated to the fields are:
\begin{equation}\label{Fbpst}
  F_{ij}^a= \frac{4\rho^2}{\left(\xi^2 + \rho^2\right)^2} \epsilon_{ija}\ , \qquad
  F_{zj}^a=\frac{4\rho^2}{\left(\xi^2 + \rho^2\right)^2}\delta_{aj}\ .
\end{equation}
The Chern-Simons action includes a term of the form
\begin{equation}\label{chernmu}
\epsilon^{MNPQ} \int\d^4x\d z\; \hA_0 F_{MN}^aF_{PQ}^a\ ,
\end{equation}
and the generated electric field is 
\begin{equation}\label{hA0flatmu}
\hA_0 = \frac{1}{8 \pi^2 a} \frac{1}{\xi} \left(1 - \frac{\rho^4}{\left(\xi^2 + \rho^2\right)^4}\right) + \mu\ ,
\end{equation}
where $\mu$ is an integration constant dual to the baryonic chemical
potential, which we set to zero.
We expand the mass equation up to order $\lambda^{-1}$ thus including
the effects of the curved background in the Yang-Mills action and of
the Chern-Simons term: The new mass formula reads \cite{Hata:2007mb}
\begin{equation}\label{masschern}
\begin{split}
M&= M  _{\rm YM}^{\rm flat}+\dM  _{\rm YM}^{} + M  _{\rm CS}\\
%&8\pi^2\kappa + \kappa \lambda^{-1}\int d^3xdz\left[-\frac{z^2}{6}\TrF_{ij}^2 + z^2 \TrF_{iz}^2\right] -\\
%& -\frac{1}{2}\kappa \lambda^{-1} \int d^3x dz \left[ (\partial_M \hA_0)^2 + \frac{1}{32\pi^2 a}\hA_0 \epsilon_{MNPQ}\Tr(F_{MN}F_{PQ}) \right] = \\
&= 8\pi^2 \kappa \left[1+\lambda^{-1}\left(\frac{\rho^2}{6}+\frac{1}{320\pi^4a^2}\frac{1}{\rho^2} + \frac{Z^2}{3}\right) \right].
\end{split}
\end{equation}
Minimizing this quantity leads us to the values of the size $\rho$: 
\begin{equation}\label{classicalrho}
\rho^2 = \frac{1}{8\pi^2 a \lambda} \sqrt{\frac{6}{5}}\ ,
\end{equation}
and the position $Z=0$ of the classical configuration \cite{Hata:2007mb}.
The scaling with an inverse power of $\lambda$ of the instanton size
justifies our expansion around flat space, since the instanton will be
localized in a small region around $Z=0$ in the large-$\lambda$
limit.
By plugging this result back into the mass equation, we find the
classical value of the mass
\begin{equation}\label{massvalue}
M= 8\pi^2 \kappa + N_c\sqrt{\frac{2}{15}}\ .
\end{equation}
The near BPST limit works because at large $\lambda$ the instanton is
very small and thus is concentrated in a region much smaller than the
radius of spacetime curvature
\cite{Hata:2007mb,Hong:2007kx,Bolognesi:2013nja}.
With the calibration \eqref{calibration}, the mass of the classical
baryon is found to be
\beq
M   \simeq M_{\rm KK}\times 1.684 \simeq 1598\GeV \ .
\eeq
There are other $1/\lambda$ corrections and also quantum $1/N_c$
corrections to be taken into account: By quantization of the moduli
space we take into account the leading quantum corrections.

Translations and rotations in flat three-dimensional space (in the
directions $x^i$) are exact symmetries.
Translations of the solution are realized by changing the parameter
$\bX$, hence they are exact moduli.
Rotations are described by an $\SO(3)$ matrix $R_{ij}$ that acts on
the coordinates as
\beq
x^i \to R^i_{\phantom{i}j}x^j\ ,
\eeq
or alternatively, by an $\SU(2)$ matrix $B$ related to $R_{ij}$ by
\beq\label{rotationSU2}
R_{ij} = \frac{1}{2}\Tr\big(\tau^i B\tau^j B^{\dag}\big).
\eeq
Rotations in flavor $\SU(2)$ space are also an exact symmetry, and are
described by the rotations of Pauli matrices $\tau^i$ with matrices
$\A$ as
\beq
\label{transftau}
\tau^i \to \A \tau^i \A^{\dag}\ .
\eeq 
Due to the particular structure of the soliton, however, the two kinds
of rotations, those in coordinate space and those in flavor space,
are related (the same that happens for Skyrmions), so that only one set
of rotational moduli are necessary to describe all possible
configurations.
This will no longer be true when we turn on the mass splitting, since
the isospin symmetry will be explicitly broken, so here we will use
only the spatial rotations as the true moduli.

The full solution is provided in Ref.~\cite{Hashimoto:2008zw}, and
since we will only care about angular velocity terms, we will neglect
the time derivatives of other moduli:
\bea\label{fullbaryon}
A_M &=& -\i f(\xi) V\left(g\partial_M g^{-1}\right)V^{-1}-\i V\de_M V^{-1}\ , \non
A_0 &=& 0\ , \non
\hA_i&=&-\frac{N_c}{16\pi^2 \kappa }\frac{\rr}{(\xr)^2}\epsilon_{iab}\chi^a x^b\ , \non
\hA_z&=&-\frac{N_c}{16\pi^2 \kappa }\frac{\rr}{(\xr)^2}\bchi\cdot\bx\ , \non
\hA_0&=&\frac{N_c}{8\pi^2 \kappa }\frac{1}{\xi^2}\left[1-\frac{\rho^4}{(\xr)^2}\right].
\eea
Inserting these fields in the action will give the moments of inertia
of the soliton.
We note that the angular velocity is denoted by $\chi^a$
here, as is standard in the Sakai-Sugimoto model \cite{Hata:2007mb}
and the $\SU(2)$-rotation matrix is here $a$, whereas in
Sec.~\ref{due} we used the notation $\Omega_i$ (see
Eq.~\eqref{eq:angular_velocity_def}) for the angular velocity 
and $A^\dag$ for the corresponding rotation matrix.
In particular, we have
\beq
\chi^a = -\i\Tr(a^\dag\dot{a}\tau^a)\ .
\eeq
The Lagrangian of the collective modes is, at the highest order in
$\lambda$, given by the instanton moduli space dynamics lifted by the
mass for the $z$-translation and size $\rho$: 
\beq
L = -M_0 + \frac{M_0}{2}\dot{\bX}^2 + \frac{M_0}{2}\dot{Z}^2- \frac{M_0\omega_Z^2}{2} Z^2 + M_0\left(\dot{\rho}^2 + \rho^2 \dot{a}_I^2\right) - M_0 \omega_{\rho}^2 \rho^2-\frac{Q}{\rho^2}\ ,
\eeq
with
\beq
M_0 = 8 \pi^2 \kappa\ ,  \qquad 
\omega_Z^2=\frac{2}{3}\ , \qquad
\omega_{\rho}^2=\frac{1}{6}\ , \qquad
Q = \frac{N_c}{40 \pi^2 a}\ .
\eeq
This is equivalent to a Skyrmion with mass $M_0$ and diagonal moment
of inertia equal to
\beq
I_0 = \frac12 M_0 \rho^2 
= \frac12 N_c \sqrt{\frac{6}{5}} \ ,
\eeq
with the calibration \eqref{calibration} yielding
\beq
I_0 \simeq  \frac{1.64}{M_{\rm KK}} \simeq 1.73 \GeV^{-1}.
\eeq
Thus, including only the classical soliton energy and the rotational
energy, we have
\begin{equation}
\begin{split}
  &M_{p,n} = M_0 + \frac{3}{8I_0} = 0.775 \GeV \ , \quad \quad
  M_{\Delta} = M_0 + \frac{15}{8I_0} = 1.64\GeV\  ,\\ 
& \qquad \qquad  \qquad  \ M_{\Delta}-M_{n,p} = \frac{3}{2I_0} = 0.87\GeV\ .
\end{split}
\label{eq:N_mass_and_Delta_mass_WSS}
\end{equation}
This is just an estimate of the values, in fact the BPST solution is
modified at large distances, larger than $1/{M_{\rm KK}}$.
This computation can be done since at those distances we are in the
linear regime so taking the curvature effect into account becomes
simpler.
The coefficient of the linear tail can be computed with the Green's
function in curved space \cite{Hashimoto:2008zw}.
The $A_z$ field in the singular gauge (neglecting the $\A$ moduli for
the moment) reads: 
\beq
A_z^{({\rm S})} = \left(\frac{1}{\xx}-\frac{1}{\xr}\right)\bx\cdot\btau\ ,
\eeq
which at distances larger than the soliton radius $(\xi\ll\rho)$, but
smaller than the curvature scale $(h(z)\simeq k(z)\simeq 1)$, is
approximated by:
\beq
A_z^{({\rm S})} \simeq   \frac{\rho^2}{\xi^4} \bx\cdot\btau = -\frac{ \rho^2}{2} \nabla\left(\frac{1}{\xi^2}\right) \cdot\btau\ .
\eeq
Note that $-\frac{1}{4 \pi^2\xi^2}$ is the Green's function in flat
$4$-dimensional space.
To generalize to the large distance region, we substitute it with
the Green's function in curved space \cite{Hashimoto:2008zw}
\beq
H(r,0,z,0) =- \kappa  \sum_{k=0}^{\infty} \phi_k(z) \phi_k(0)\frac{e^{-\sqrt{\lambda_k} r}}{4\pi r}\ ,
\eeq 
where we already set the semiclassical value $Z=0$; we chose
$\bX=0$ without loss of generality and inserted
the mesonic eigenmodes and eigenvalues defined as solutions of the
Sturm-Liouville equation:
\beq
-h(z)^{-1} \partial_z \left(k(z) \partial_z \psi_n(z)\right) = \lambda_n \psi_n(z)\ .
\eeq
The $n=0$ solution corresponds to a non-normalizable mode with zero
mass (before taking into account the Aharony-Kutasov action), dual to
the pion wave function, while the other $n>0$ modes correspond to
massive vector mesons:
\begin{align}
  \label{eq:pionphi}
  \phi_0(z) &= \frac{1}{\sqrt{\kappa \pi}} \frac{1}{k(z)}\ ,\\
  \phi_n(z) &= \frac{1}{\sqrt{\lambda_n}}  \partial_z  \psi_n(z)\ .
\end{align}
These modes (except the $n=0$ one) can be normalized with the relations
\beq
\kappa \int\d z\; h(z) \psi_n(z) \psi_m(z) = \delta_{nm}\ ,\qquad
\kappa \int\d z\; k(z) \phi_n(z) \phi_m(z) = \delta_{nm}\ .
\eeq
The Skyrme truncation amounts to including only the pion (and the
$\eta$, that shares the same mode), so that the $A_z^{({\rm S})}$
field is expanded as:
\beq
A_z^{({\rm S})} \simeq  -\frac{ \pi}{2} \kappa \rho^2 \phi_0(z) \phi_0(0) \nabla\left(\frac{1}{r}\right)\cdot\btau + \sum_{n=1}^{\infty} \cdots
\eeq 
In order to read off the coefficient of the 3-dimensional linear tail,
we need to isolate the $r$-dependence, while integrating the pion
profile in the holographic direction.
Using Eq.~\eqref{eq:pionphi} we obtain
\beq
\int_{-\infty}^{+\infty} \d z\; A_z^{({\rm S})} \simeq   -\pi \kappa \rho^2\frac{1}{2\kappa \pi} \int_{-\infty}^{+\infty} \d z\; \frac{1}{k(z)}\nabla\left(\frac{1}{r}\right) \cdot\btau\ .
\eeq 
The integral gives a factor of $\pi$ and we can use the semiclassical size
$\rho^2 = \frac{N_c}{8\pi^2\kappa}\sqrt{\frac{6}{5}}$ to obtain
\beq
\int_{-\infty}^{+\infty} \d z\; A_z^{({\rm S})} \simeq  -\frac{N_c}{16\pi\kappa}\sqrt{\frac{6}{5}}\nabla\left(\frac{1}{r}\right) \cdot\btau\ .
\label{eq:Ctail_WSS}
\eeq
Defining the coefficients of the tail of the non-rotating soliton (in
the standard orientation of the hedgehog) as
\begin{align}
\frac12\int \d z\; A_z^{a=1,2} &= -C \p_{1,2}\left(\frac{1}{r}\right)
\ ,\\
\frac12\int \d z\; A_z^{a=3} &= -\frac12(D+E)\p_{3}\left(\frac{1}{r}\right)\ , \\
\frac12\int \d z\; \widehat{A}_z &= -\frac12(D-E)\p_{3}\left(\frac{1}{r}\right)\ .
\end{align}
The final result is then read off of Eq.~\eqref{eq:Ctail_WSS} as
\beq
C=D=E= \frac{N_c}{16\pi\kappa}  \sqrt{\frac{6}{5}} =   c_0 \frac{1}{\lambda}\ ,
\eeq
using that $\widehat{A}_z=0$ for the non-rotating soliton.
The coefficient $c_0$ is thus
\beq
c_0 =  \frac{1}{16\pi a}  \sqrt{\frac{6}{5}}  =\frac{27\pi^2}{2}\sqrt{\frac{6}{5}}\ .
\eeq
This is exact in the large-$\lambda$ limit.
To compare it with the coefficient found for the Skyrmion, we need to
recall that we are using different units in the two computations; here
we adopted dimensionless units, meaning that dimensionful quantities
need to have factors of $M_{\rm KK}$ restored. The integral of the
gauge field is dimensionless, so the coefficient $C$ must have the
dimension of $M_{\rm KK}^{-2}$ to cancel that of $\nabla\frac{1}{r}$:
\beq
C^{\rm dim} = C M_{\rm KK}^{-2} =\frac{27 \pi^2}{2M_{\rm KK}^2\lambda}\sqrt{\frac{6}{5}} =9.75\times 10^{-6}\MeV^{-2}=(0.61\fm)^2\ .
\eeq

The introduction of explicit breaking of isospin symmetry within the
holographic model of Sakai-Sugimoto is realized by changing the mass
matrix in Eq.~\eqref{SAK} to something not proportional to the
identity matrix, resulting in the moment of inertia being modified,
taking the form of Eq.~\eqref{intenssplit}.
 
Let's consider the baryon masses at $\epsilon=0$ and how they are
affected by the quark mass.
Various effects have been considered in the past, e.g.~the leading-order effect for $\epsilon=0$ and a finite quark mass, is a shift in
the meson and baryon spectra in both the 2-flavor
\cite{Hashimoto:2009hj} and 3-flavor \cite{Hashimoto:2009st} cases.
Here, we will add also the modification to the inertia which in turn
affects the nucleon-Delta splitting.   
The formula \eqref{intenssplit} unfortunately cannot yet be used
reliably at this stage to compute the mass correction to the inertia
$I$.
The point is that we know $I_0$ from the selfdual instanton
approximation, and we know that it receives, even at $m=0$, order
$1/\lambda$ corrections from the curvature effects.
The mass corrections are of the same order $\kappa C^2 \propto 1/\lambda$.
So to make use of Eq.~\eqref{intenssplit}, we should first find a
technique to compute also $I$ up to order $1/\lambda$. 

However, the splitting term $\delta$ can be computed within the
approximation of small $m$: This is possible because the dominant
contribution to $\delta$ in this limit comes from the soliton tails,
which can be computed exactly.
The linearized tail is a reliable approximation from a certain cutoff
$r=R_{\rm UV}$ up to $r\rightarrow\infty$: From the WSS model we know
that the cutoff should be roughly of the order of the soliton size
$R_{\rm UV}\sim \rho$.
As we will see, and as has happened in the Skyrmion case, employing
the tail to compute the moment of inertia will lead to a divergent
result, a hint that the correct configuration for $r<R_{\rm UV}$ is
not given by the linearized tail, but by some perturbation of the BPST
instanton.
However, the divergent term is independent of $m$, while the leading
order in $m$ is linear: This linear term does not depend on $R_{\rm UV}$,
allowing us to compute the splitting explicitly.
Performing the full numerical solution is more challenging as compared
to the Skyrme model, due to the presence of the additional holographic
coordinate.
On the other hand in the WSS model we have an analytic approximation
in the BPST instanton for the core configuration, and a semi-analytical
expression as an expansion over mesonic modes for the linearized tail:
We now move to compute the deformation of the tail induced by the
presence of $S_{\rm AK}$.

As a first step, we write the linearized equations of motion: Only the
fields $\widehat{A}_z$ and $A_z^{a=3}$ will take part in the splitting,
hence we have
\beq
\kappa k(z)\left( \partial_i^2 \widehat{A}_z -  \partial_i \partial_z \widehat{A}_i \right) &=& m_qc\left[ \int_{-\infty}^{+\infty}\d z\;\widehat{A}_z + \epsilon \int_{-\infty}^{+\infty}\d z\; A_z^{a=3} \right]\ ,\\
\kappa k(z)\left( \partial_i^2 A_z^{a=3}-  \partial_i \partial_z A_i^{a=3} \right) &=& m_qc\left[ \int_{-\infty}^{+\infty}\d z\;A_z^{a=3} + \epsilon \int_{-\infty}^{+\infty}\d z\;\widehat{A}_z \right]\ .
\eeq
We immediately see that the equations are coupled: This is a consequence of the mass matrix of the pions (and $\eta$) becoming nondiagonal. It is simple to diagonalize the mass matrix, or analogously to diagonalize these two equations. To do so we introduce the mass eigenstates
\beq
A_\eta =\frac{1}{\sqrt{2}}\big(A_z^{a=3}+\widehat{A}_z\big)\ ,\qquad
A_\pi = \frac{1}{\sqrt{2}}\big(A_z^{a=3}-\widehat{A}_z\big)\ .
\eeq
We can in principle perform the change of basis for the fields also
for the spatial components $A_i$, but it is not necessary: The
Aharony-Kutasov action, in fact, introduces effective mass terms only
for the pseudoscalar degrees of freedom, as can be easily understood
once we expand the field $A_z$ in mesonic modes.
Then the holonomy of the field $A_z$ can be written as 
\beq
\int_{-\infty}^{+\infty}\d z\; A_z = \sum_{n=0}^{\infty} \varphi_n(x)\int_{-\infty}^{+\infty}\d z\;\phi_n(z)= \phi_0(x)\pi +  \sum_{n=1}^{\infty}  \varphi_n(x)\int_{-\infty}^{+\infty}\d z\;\partial_z\psi_n(z)\ ,
\eeq
and since the functions $\psi_n(z)$ vanish at the UV boundary, only
the first term in the sum survives, generating a mass term for the
pseudoscalars.
Because of this, we will restrict our analysis to the $z$ component of
the fields: After diagonalization it is easy to read the mass
eigenvalues as being
\beq
m_\pm^2= m^2 \left(1\pm \epsilon\right),
\eeq
so we can make an educated guess for the shape of the solution: The
asymptotic configuration in the massless scenario is given by
\beq\label{eq:asymptoticwhAz}
\widehat{A}_z &\simeq& \frac{\rho^2 N_c}{8 \kappa}\chi^j \partial_{X^j} H(\bx,\bX,z,Z)\ ,\\
A_z^{a=3} &\simeq& -2\pi^2\rho^2\Tr\left(a \tau^k a^{\dagger}\tau^3\right)\partial_{X^k}H(\bx,\bX,z,Z)\ ,\label{eq:asymptoticAz3}
\eeq
with
\beq
H(\bx,\bX,z,Z) = \kappa \sum_{n=0}^\infty\phi_n(z)\phi_n(Z)Y_n(r)\ ,\qquad
Y_n(r) = -\frac{1}{4\pi}\frac{e^{-\sqrt{\lambda_n}r}}{r}\ ,
\eeq
and we can easily think of keeping this general shape while providing
a nonvanishing $\lambda_0$ to account for the pion mass.
However, we have a mass splitting here, so we need to introduce two
values $\lambda_\pm=m^2 \left(1\pm \epsilon\right)$.
Moreover, the mass eigenstates are combinations of the fields
\eqref{eq:asymptoticwhAz},\eqref{eq:asymptoticAz3}, so we build an
Ansatz for $A_\eta$, $A_\pi$ given by
\beq
A_\eta&=& b\chi^j\partial_{X^j}H_+ - \tilde{b} \Tr \left(a\tau^j a^\dagger \tau^3\right)\partial_{X^j}H_+\ ,\label{eq:Aeta}\\
A_\pi&=& d\chi^j\partial_{X^j}H_- - \tilde{d} \Tr \left(a\tau^j a^\dagger \tau^3\right)\partial_{X^j}H_-\ .\label{eq:Api}
\eeq
The constants should be determined by the boundary conditions (the
matching with the core configuration in an intermediate region), while
the functions $H_{\pm}$ are given by
\beq
H_{\pm}\equiv -\kappa \sum_{n=0}\phi_n(z)\phi_n (Z)\frac{1}{4\pi}\frac{e^{-\sqrt{\lambda_n}r}}{r}\ ,\qquad
\lambda_0 = m_\pm^2\ .
\eeq
With this choice, the equations of motion are satisfied, and we only need to fix the constants: We then require that in the massless limit the fields revert to
\beq
\lim_{m\to0}A_\eta =\widehat{A}_z\ ,\qquad
\lim_{m\to0}A_\pi =A_z^{a=3}\ ,
\eeq
and we obtain 
\beq
b=-d\ ,\qquad
\tilde{b}=\tilde{d}\ .
\label{eq:bbtilde}
\eeq
To conclude, we also require matching with the core of the baryon
solution, which is equivalent to requiring that the coefficients of
Eqs.~\eqref{eq:asymptoticwhAz} and \eqref{eq:asymptoticAz3} are
reproduced.
This fixes all the constants as
\beq
b= \frac{\rho^2 N_c}{\sqrt{2}8\kappa}\ ,\qquad
\tilde{b} = \sqrt{2}\pi^2 \rho^2 = C=D=E\ .
\eeq 
Now that the solution is fixed, we note that since only the neutral
pion and $\eta$ masses are modified, we can neglect the vector mesons in
the computation of the splitting: They will contribute to the total
moment of inertia, but with a term proportional to the identity
matrix. To isolate the pseudoscalars, we remove them from the sum in
the functions $H_\pm$ and define new Yukawa potentials $Y_\pm(r)$:
\beq
H_{\pm}=\kappa \phi_0(z)\phi_0 (Z)Y_\pm(r) +\kappa \sum_{n=1}^\infty\phi_n(z)\phi_n (Z)Y_n(r)\ ,\qquad Y_\pm(r)=-\frac{1}{4\pi}\frac{e^{-m_\pm r}}{r}\ .
\eeq
We can read off the inertia tensor from the terms in the Lagrangian
that are quadratic in the angular velocity $\bchi$: Such contributions
can arise from all terms of the action, the Yang-Mills, Chern-Simons
and Aharony-Kutasov parts.
The leading contribution to the inertia splitting is given in the
small-$m$ limit by a linear term, hence we discard corrections coming
from the Aharony-Kutasov term, whose prefactor is of order
$m_q c\sim m^2 f_\pi^2$.
Moreover, the leading contribution comes from the linearized soliton
tail, hence we neglect the Chern-Simons term, being at least cubic in
the fields.
We are then left with the Yang-Mills action to be computed on the
solution of the equations of motion just obtained.
The action terms we are looking for are the ones involving the
$\mathcal{A}_z$ and at most quadratic in the fields:
\beq
S_{\rm YM}|_{\mathcal{A}_z} = -\kappa \Tr\int \d^4x\d z \; k(z)\left[\left(\partial_i\mathcal{A}_z\right)^2 -\left(\partial_0\mathcal{A}_z\right)^2\right],
\eeq
which receives contributions both from the $A_z^{a=1,2}$ and the
$A_\eta$, $A_\pi$ fields,
which we split into two terms, defining the kinetic energy parts as
\begin{align}
  T^{(1,2)} &= \frac{\kappa}{2}\sum_{a=1}^2\int\d z\d^3x\; k(z)\left[\left(\partial_iA_z^{a}\right)^2 -\left(\partial_0A_z^{a}\right)^2\right],\non
T^{(3)} &= \frac{\kappa}{2}\int\d z\d^3x\; k(z)\left[\left(\partial_iA_z^{3}\right)^2 -\left(\partial_0A_z^{3}\right)^2+\big(\partial_i\widehat{A}_z\big)^2 -\big(\partial_0\widehat{A}_z\big)^2\right]\non
&=\frac{\kappa}{2}\int\d z\d^3x\; k(z)\left[\left(\partial_iA_\eta\right)^2 -\left(\partial_0A_\eta\right)^2+\big(\partial_iA_\pi\big)^2 -\big(\partial_0A_\pi\big)^2\right],
\end{align}
where we have used Eq.~\eqref{notationsu2} 
and the total kinetic energy is $T=T^{(1,2)}+T^{(3)}$.
Starting from the latter, which gives rise to the following terms at
order $\chi^2$:
\begin{align}
  \label{eq:DiAzterms}
  &\Tr\int \d z\d^3x\; k(z)\left[\left(\partial_iA_\eta\right)^2 +\left(\partial_iA_\pi\right)^2\right]= \non
  &\qquad b^2\chi^j\chi^k\int\d^3x\d z\; k(z)\left(\partial_i\partial_jH_+\partial_i\partial_k H_++\partial_i\partial_jH_-\partial_i\partial_k H_-\right),\\
\label{eq:D0Azterms}
&-\Tr\int \d z\d^3x\; k(z)\left[\left(\partial_0A_\eta\right)^2 +\left(\partial_0A_\pi\right)^2\right]= \non
&\qquad -4\tilde{b}^2\dot{R}_{a3}\dot{R}_{k3}\int \d^3x\d z\; k(z)\left(\partial_aH_+\partial_k H_++\partial_aH_-\partial_k H_-\right).
\end{align}
We see immediately that the terms in Eq.~\eqref{eq:DiAzterms} produce
kinetic energy contributions proportional to the identity matrix,
since after integrating over the angular coordinates of $\mathbb{R}^3$
we are left with something proportional to $\bchi\cdot\bchi$.
On the other hand, the terms in Eq.~\eqref{eq:D0Azterms} do break
spherical symmetry, since they select the direction in coordinate
space that corresponds to the rotation performed by $R_{a3}$, that is,
the direction that we obtain for the orientation of $\pi^3$ after the
moduli $\A$ are applied to the soliton configuration: With the
standard hedgehog orientation, this direction would simply be given by
$\hat{x}^3$.
We then introduce the body-fixed axis and give the components of the
angular velocity in terms of this coordinate system: We label them
$(\chi_\xi, \chi_\eta,\chi_\zeta)$, where $\chi_\zeta$ identifies the
projection along the axis of residual symmetry (again, in the case of
standard orientation, $\chi_\zeta=\chi_3$).
This leads us to consider only the terms in Eq.~\eqref{eq:D0Azterms},
which can then be computed by using the relations:
\beq
\dot{R}_{ab} &=&\epsilon_{amn}\chi_mR_{nb}\ ,\\
\int \d^3x\; \partial_i f(r) \partial_j f(r) &=& \frac{4\pi}{3} \delta_{ij}\int \d r\; r^2 f'(r)^2\ ,\\
\epsilon^{akb}\epsilon^{acd}\chi^k\chi^c R_{b3}R_{d3} &=& \chi^2 R_{b3}R_{b3} -\left(\chi^aR_{a3}\chi^bR_{b3}\right)=\chi^2-\chi_\zeta^2\ ,\label{eq:asym3}\\
\int_R^{\infty}\d r\;\frac{\left(m_\pm r+1\right)^2}{r^2}e^{-2m_\pm r} &=& \left(\frac{m_\pm}{2}+\frac{1}{R}\right)e^{-2m_\pm R}\simeq \frac{1}{R}-\frac{3}{2}m_\pm + \mathcal{O}(m^2)\ .
\eeq
After keeping only the $n=0$ term in the mesonic modes' expansion of
the fields (higher $n$ does not contribute to the splitting, while
cross terms vanish because of orthogonality relations) we can write
the following kinetic energy term $T^{(3)}$ for the fields
$A_\eta,A_\pi$:\footnote{The mass shift induced on the configuration
by including the quark mass for $\epsilon=0$
\cite{Hashimoto:2009hj,Hashimoto:2009st} is an order $m^2$ effect and
alters only the potential energy.}
\beq
T^{(3)}= \frac13\pi^2\kappa\rho^4\left(\chi^2-\chi_{\zeta}^2\right)\left[\frac{2}{R}-\frac{3}{2}\left(m_++m_-\right)\right] + \mathcal{O}(m^2)\ .
\eeq
This formula presents a UV divergent term proportional to $R^{-1}$,
with $R$ being a cutoff to the integral over $r$, which arises from
the extrapolation of the linear tail to the core region.
Realistically, the cutoff $R$ is not to be sent to zero, but rather to
some value of the order of the soliton size $\rho$.
In this way, we can provide an estimate of the order of this term,
however, it is not necessary since we can argue that it cannot
contribute to the splitting: In fact, this term is independent of the
parameters $m,\epsilon$, so we cannot regard it as a correction to the
inertia due to the presence of the Aharony-Kutasov term.
It is instead to be interpreted as the correction to the inertia due
to the deviation of the tail of the soliton from the BPST instanton
configuration, ultimately due to the curved background.

The last part we need to compute is the contribution to the inertia of
the fields $A_z^{a=1,2}$: The computation is analogous to the one
performed above, with the only exception being that every instance of
$m_\pm$ is substituted with $m$, and formula \eqref{eq:asym3} has to
be modified in favor of the components $\chi_\xi,\chi_\eta$. 
The resulting contribution is then given by
\beq
T^{(1,2)} = \frac{2}{3}\pi^2\kappa\rho^4\left(\chi^2+\chi^2_\zeta\right)\left[\frac{1}{R}-\frac{3}{2}m\right] + \mathcal{O}(m^2)\ .
\eeq
The sum of the two terms finally gives the complete rotational kinetic energy $T=T^{(1,2)}+T^{(3)}$:
\begin{equation}
T=\frac{\pi^2\rho^4\kappa}{2}m\left[\left(\frac{8}{3mR}-\left(2+\sqrt{1+\epsilon}+\sqrt{1-\epsilon}\right)\right)\chi^2 - \left(2-\sqrt{1+\epsilon}-\sqrt{1-\epsilon}\right)\chi^2_\zeta\right].
\end{equation}
We thus confirm that the UV divergent term does not enter in the
splitting, and we can see by expanding the square roots that the
splitting term, $\delta$, appears at order $m\epsilon^2$.
We can then compute the splitting $\delta$ of Eq.~\eqref{intenssplit}:
\beq
\delta = \frac{\pi^2 \rho^4\kappa}{12}m\epsilon^2 +
\mathcal{O}(m^2,\epsilon^4)\ .
\eeq

\subsection{New quantization for \texorpdfstring{$\epsilon\neq 0$}{epsilon=\=0} and \texorpdfstring{$\Delta$}{Delta} splitting}
\label{sec:newquant}

In this section, we derive a quantum mechanical Hamiltonian for the
baryon spectrum up to order $\epsilon^2$:
To do so, we take into account two main effects: One is the
perturbation of the moment of inertia due to the presence of the quark
mass term, which we already analyzed in the small-$m$ limit, while the
second is the presence of a term linear in the angular velocity in the
action, arising from the Aharony-Kutasov action evaluated on the
unperturbed baryon configuration (it arises in general on the full
configuration, and we can see the presence of linear terms already in the
previous section, arising from cross terms proportional to
$b\tilde{b}$ or $d\tilde{d}$ (see Eq.~\eqref{eq:bbtilde} and the
Ans\"atze \eqref{eq:Aeta}-\eqref{eq:Api}),
but they are subleading contributions in $m$ when
compared to the ones arising from the unperturbed baryon core).

We have already computed the new moments of inertia of the soliton, so
the kinetic part of the Lagrangian coming from the Yang-Mills action,
neglecting the motion of the center of mass, reads: 
\beq
L_{\rm YM} = \frac{1}{2}I_A \left(\chi_{\xi}^2+\chi_{\eta}^2\right)+\frac{1}{2}I_C \chi_{\zeta}^2 -M\ ,
\eeq 
where
\beq
I_A = I +\delta \ , \qquad I_C = I -2 \delta\ .
\eeq
Again, we have relaxed the choice of a frame in which the axis of the
cylindrical symmetry is oriented along $\hat{x}^3$: We decomposed the
angular velocity in components along principal axes of inertia of this
``rigid top'', while the spatial orientation of these axes is
completely free.

If this was the complete Lagrangian, it would be straightforward to
get the Hamiltonian: We simply need to trade angular velocities for
their conjugate momenta $J_j=\calI_{jk}\chi_k$ to obtain the Hamiltonian
of a rigid symmetrical top as in Eq.~\eqref{eq:H_rot}:
\beq
E_{\rm top} =  \frac{1}{2I_A}j(j+1)+\frac{1}{2}\left(\frac{1}{I_C}-\frac{1}{I_A}\right) j_3^2 +M\ .
\label{symtop}
\eeq
However, it would be naive to extract the Hamiltonian by simply
identifying $J_j=\calI_{jk}\chi_k$:
The Aharony-Kutasov action provides a Lagrangian term linear in the
angular velocity. This term arises at the first order in perturbation
theory, hence is obtained by feeding the unperturbed baryon
configuration to the perturbation Lagrangian, in this case provided by
the Aharony-Kutasov term \eqref{SAK}. As noted before, at the static
order ($\bchi=0$) the terms proportional to $\epsilon$ vanish due
to the trace: The first nonvanishing $\mathcal{O}(\epsilon)$ term arises at
linear order in $\chi$, hence is an $N_c^{-1}$ correction to the baryon
mass.
To show the mechanism as presented in Ref.~\cite{Bigazzi:2018cpg}, we start
with the Aharony-Kutasov Lagrangian written in terms of $U$, $\varphi$ as
defined in Eq.~\eqref{AzU}:
\beq
L_{\rm AK} = c\Tr\int\d^3x\left[Me^{\i\varphi}\A U\A^\dag + e^{-\i\varphi}\A U^\dag\A^\dag M -2\mathbb{1}_2\right].
\eeq
We then write the matrix $U$ using the BPST instanton approximation:
\beq
U= -\cos\alpha + \i \sin \alpha \;\hat{\bx}\cdot\btau\ ,\qquad
\alpha = \frac{\pi}{\sqrt{1+\frac{\rho^2}{r^2}}}\ ,
\eeq
where the function $\alpha(r)$ is defined by the integral over $z$ of
the field $A_z$ in the unperturbed baryon configuration.
We can then keep the terms proportional to $\epsilon$ and expand the
exponential to linear order in $\varphi$:
\beq
L_{\rm AK}^{\epsilon} &=& 2m_qc\epsilon \int \d^3x\; \varphi \sin\alpha \frac{x^i}{r} \Tr \left[\tau^3\A \tau^i
 \A^\dag\right]\non
 &=& \frac{m_qc\epsilon N_c}{16\pi\kappa} \int \d^3x\; \sin\alpha \frac{1}{\rho\left(1+\frac{r^2}{\rho^2}\right)^{\frac{3}{2}}} \frac{x^ix^j}{r} R_{3i}\chi^j\non
 &=& \frac{m_qc\epsilon N_c}{12\kappa}\rho^3 \mathcal{J}_2 \chi^i R_{3i}\ ,
\eeq
where we made use of the $\widehat{A}_z$ field as in Eq.~\eqref{fullbaryon} and we defined the integral
\beq
 \mathcal{J}_2 \equiv \int_{0}^{\infty}\d y\;\frac{y^3}{\left(1+y^2\right)^{\frac{3}{2}}}\sin\left(\frac{\pi}{\sqrt{1+y^{-2}}}\right)\simeq 1.054\ .
\eeq
The term $L_{\rm AK}^\epsilon$ modifies the canonical relation between angular velocity and the angular momentum, which is now given by
\beq
J_i = \calI_{ij} \chi_j +  \frac{\epsilon m_q c N_c}{12\kappa}\rho^3 R_{3i} \mathcal{J}_2
=\calI_{ij} \chi_j + \frac{\epsilon m^2 N_c}{12\pi}\rho^3 R_{3i} \mathcal{J}_2\equiv \calI_{ij}\chi_j -K_i\ .
\eeq
Notice that in the definition of $K_i$ there appears the rotation
matrix $R_{3i}$. This means that the vector $K_i$ always points in the
direction labeled by $\zeta$ (that is, the direction that in the
body-fixed frame becomes $\hat{x}_3$). This implies that only the relation
between $J_{\zeta}$ and $\chi_{\zeta}$ is modified, and the full
Hamiltonian is thus given by 
\beq
H_{\rm full} &=& \frac{1}{2I_A} \left(J_{\xi}^2+J_{\eta}^2\right)+\frac{1}{2I_C} \left(J_{\zeta}^2+K_{\zeta}^2+2J_{\zeta}K_{\zeta}\right) +M_0\non
&=&  \frac{1}{2I_A}J^2+\frac{1}{2}\left(\frac{1}{I_C}-\frac{1}{I_A}\right) J_{\zeta}^2+ \frac{1}{I_C}J_{\zeta}K_{\zeta}+\frac{1}{2I_C}K_{\zeta}^2+M\ ,
\eeq
where we have defined $K_{\zeta}\equiv-\frac{\epsilon m_q c N_c}{12\kappa}\rho^3\mathcal{J}_2$.
We see that the Hamiltonian is still diagonal and the energies are
simply obtained as 
\beq
E_{\rm full} =  \frac{1}{2I_A}j(j+1)+\frac{1}{2}\left(\frac{1}{I_C}-\frac{1}{I_A}\right) j_3^2+ \frac{1}{I_C}K_{\zeta}j_3+\frac{1}{2I_C}K_{\zeta}^2 +M\ .
\eeq
We may then use the relation $J^iR_{ki} = -I^k $ to identify
$J_{\zeta}\equiv -I_3$  so that the quantum number $j_3$ can be replaced
by $-i_3$. We also note that the corrections $\Delta I$ are of order
$\epsilon^2$, while $K_{\zeta}$ is of order $\epsilon$: If we stop our
analysis at order $\epsilon^2$ we can safely replace $I_C$ with the
unperturbed $I_0$ in the terms proportional to $K_{\zeta}$, leading to
\beq
E_{\rm full} =  \frac{1}{2I_A}j(j+1)+\frac{1}{2}\left(\frac{1}{I_C}-\frac{1}{I_A}\right) i_3^2- \frac{1}{I_0}K_{\zeta}i_3+\frac{1}{2I_0}K_{\zeta}^2 +M\ ,
\eeq
We observe two new terms in the energy formula with respect to
Eq.~\eqref{symtop}. $\frac{1}{2I_0}K_{\zeta}^2$ is simply a mass shift
due to the quark mass difference: We see that it is always positive
and of order $\epsilon^2 m$.  $-\frac{1}{I_0}K_{\zeta}i_3$ is the
isospin-breaking term that is responsible for the splitting of states
with the same absolute value of $i_3$, i.e.~the term computed in
Ref.~\cite{Bigazzi:2018cpg} that accounts for the proton-neutron mass
difference. It also removes half of the degeneracy in the $\Delta$
multiplet. It is of order $\epsilon m^2$.

\begin{figure}[!ht]
  \centering
  \includegraphics[width=0.7\textwidth]{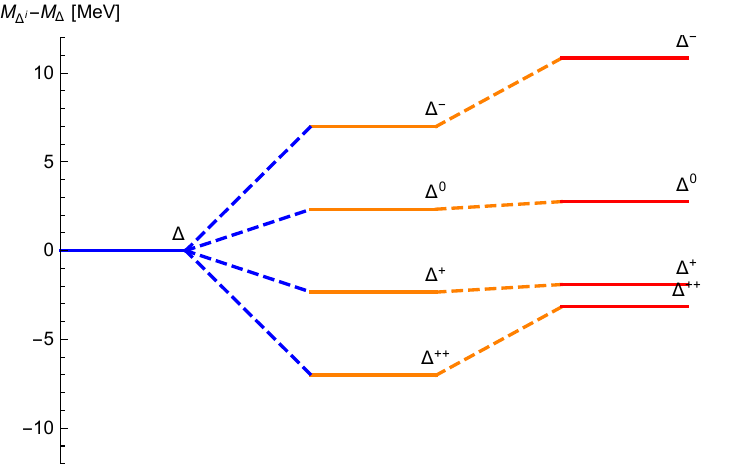}
  \caption{
    The mass spectrum of the Delta multiplet, measured with
    respect to the semi-classical mass.
    The first splitting (from left) shown with blue-dashed lines is due to the
    splitting that is linear in the isospin quantum number $j_3$, but
    quadratic in the pion mass $m$, corresponding to the result of
    Ref.~\cite{Bigazzi:2018cpg}.
    The orange-dashed correction shown subsequently is our result that
    is quadratic in $j_3$, but linear in $m$.
    }
  \label{fig:DeltaLevels}
\end{figure}

\begin{figure}[!ht]
  \centering
  \includegraphics[width=0.7\textwidth]{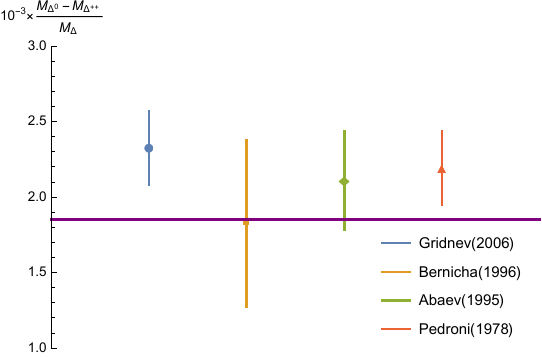}
  \caption{
    The mass splitting between the $\Delta^{++}$ state and the
    $\Delta^0$ state, which is measured experimentally by different
    groups: Pedroni et.al.~\cite{Pedroni:1978it}, Abaev
    et.al.~\cite{Abaev:1995cx}, Bernicha
    et.al.~\cite{Bernicha:1995gg} and Gridnev
    et.al.~\cite{Gridnev:2004mk}.
    Our result is shown with the purple horizontal line.
  }
  \label{fig:DeltaSplitPlot}
\end{figure}

We immediately see that there are two contributions to the splitting
of the $\Delta$ multiplet: Terms of order $\epsilon$ (giving higher
mass to lower values of $i_3$) and terms of order $\epsilon^2$ (not
depending on the sign of isospin, but on the absolute value). This is
exactly what is suggested by the separation geometry model (see
Ref.~\cite{Filewood:2002hr}).
The estimate we obtained for the splitting of the moment of inertia is
consistent only in the $m\ll|\epsilon|$ limit: We need to keep this in
mind, as we have to sum terms of order $m\epsilon^2$ (from $\delta$)
and terms of order $m^2\epsilon$ (from $K_\zeta$).
Then we consider the leading-order term in the splitting to be of
order $m\epsilon^2$, which gives
\beq
\frac{1}{2} (M_{\Delta^{++}} - M_{\Delta^{+}}  -M_{\Delta^{0}} + M_{\Delta^{-}}) =
\frac{3}{2}\frac{\delta}{(I+\delta)(I - 2 \delta)}  2 \simeq   \frac{3\delta }{I_0^2}\ .
\label{prolatesplitting}
\eeq
Assuming the previous results we get
\beq
\frac{1}{2} (M_{\Delta^{++}} - M_{\Delta^{+}}  -M_{\Delta^{0}} + M_{\Delta^{-}}
)  \simeq  \frac{\pi^2 \rho^4\kappa}{4I_0^2}m\epsilon^2 
\simeq\frac{27\pi}{8}\frac{m\epsilon^2}{N_c\lambda}\ .
\eeq
For the numerical values we obtain
\beq
\frac{1}{2} (M_{\Delta^{++}} - M_{\Delta^{+}}  -M_{\Delta^{0}} + M_{\Delta^{-}}
)   \simeq 3.41\MeV \ .
\eeq
The splitting of order $\epsilon m^2$, which splits even proton and
neutron masses, is instead given by 
\beq
M_{p} - M_{n} = \frac{\epsilon m^2 N_c}{48\pi^3\kappa}\rho\mathcal{J}_2\simeq - 4.74\MeV \ .
\label{CSsplitting}
\eeq
A dimensionless quantity, we can compute, is the ratio of the splitting
with the semi-classical nucleon mass \cite{ParticleDataGroup:2020ssz}
\beq
\frac{M_{p} - M_{n}}{M_{N}} = -1.78\times 10^{-3}, \qquad
\left[\frac{M_{p}-M_{n}}{M_{N}}\right]_{\rm exp} = -1.38\times 10^{-3}\ ,
\eeq
where we have used the quantum-corrected approximate mass \cite{Hata:2007mb}:
\beq
M(\ell) = 8\pi^2\kappa + \sqrt{\frac{(\ell+1)^2}{6}+\frac{2N_c^2}{15}} + \frac{2}{\sqrt{6}}\ ,
\label{eq:MN_Hata}
\eeq
with the nucleon mass $M_N=M(1)$.
The full pattern of splittings in the $\Delta$ multiplet is then
obtained by combining these two contributions: 
\begin{align}
M_{\Delta^{++}}-M_{\Delta^+} &= \frac{3\delta}{I_0^2} +\frac{\epsilon m^2 N_c}{48\pi^3\kappa}\rho\mathcal{J}_2 = -1.25 \MeV\ , \nn\\
M_{\Delta^{+}}-M_{\Delta^0} &= \frac{\epsilon m^2 N_c}{48\pi^3\kappa}\rho\mathcal{J}_2 = -4.67 \MeV\ ,\nn \\
M_{\Delta^{0}}-M_{\Delta^-} &= - \frac{3\delta}{I_0^2} +  \frac{\epsilon m^2 N_c}{48\pi^3\kappa}\rho\mathcal{J}_2 = -8.08\MeV\ .
\end{align}
The ratio of the splitting with the semi-classical Delta mass is
\begin{align}
  \frac{M_{\Delta^{++}}-M_{\Delta^+}}{M_{\Delta}}&= -3.92\times 10^{-4}\ ,\non
  \frac{M_{\Delta^{+}}-M_{\Delta^0}}{M_{\Delta}}&= -1.46\times 10^{-3}\ ,\non
  \frac{M_{\Delta^{0}}-M_{\Delta^-}}{M_{\Delta}}&= -2.53\times 10^{-3}\ ,
  \label{eq:DeltaSplittings}
\end{align}
with the Delta mass $M_\Delta=M(3)$ of Eq.~\eqref{eq:MN_Hata}.
The mass splitting is illustrated in Fig.~\ref{fig:DeltaLevels}, where
we show the two consecutive splittings with different colors.
The first splitting shown with blue-dashed lines is due to the
splitting that is linear in the isospin quantum number $j_3$, but
quadratic in the pion mass $m$, corresponding to the result of
Ref.~\cite{Bigazzi:2018cpg}.
The orange-dashed correction shown to the right, is our result of the
splitting that is quadratic in $j_3$, but linear in $m$.
Notice that our splitting contribution does not distinguish the sign
of the isospin quantum number, but only the magnitude.
It therefore only shifts the neutron-proton masses upwards without
contributing to the neutron-proton mass difference. 
The sum of the first two splittings in Eq.~\eqref{eq:DeltaSplittings}
is measured experimentally
\cite{Pedroni:1978it,Abaev:1995cx,Bernicha:1995gg,Gridnev:2004mk} and
compared with our result in Fig.~\ref{fig:DeltaSplitPlot}.

\section{Conclusion and discussion}
\label{conclusion}

We summarize the main novel aspects presented in this paper:
\begin{itemize}
\item We have developed a technique to study the first-order
  contribution to the moments of inertia due to the mass term. This
  term is due to the linear tail effect and is linear in the pion
  mass.
\item We used this technique to compute the deviation of the inertia
  tensor of the Skyrmion, also in the presence of isospin breaking due
  to the quark mass difference. In this case, we could also perform
  the numerical computation for all values of the mass, and we
  confirmed that the technique works well when the mass is small.
  This allows us to compute a partial splitting in the baryon states,
  for example in the $\Delta$ spectrum, but it is not able to split the
  neutron-proton masses. 
\item We then performed the computation of the splitting of the
  moments of inertia in the WSS model of holographic QCD. Here we
  cannot perform the numerical computation due to the complexity of the
  problem, but we can apply our technique in the small-mass limit. 
\item In the WSS model, there is also another source of splitting of
  baryon masses due to the presence of vector mesons and the
  Chern-Simons term \cite{Bigazzi:2018cpg} (this mass splitting
  mechanism is similar to that of Ref.~\cite{Jain:1989kn} in the
  context of the Skyrme model with $\eta$, $\omega$ and $\rho$ mesons).
  Here we can thus see the combination of these effects, with the one
  we derived splitting the moments of inertia.
  This allowed us to complete the
  spectrum of the $\Delta$ multiplet with no unwanted degeneracy and
  no equidistant splitting between the states.
\end{itemize}

We computed the mass splitting of the Deltas in the Skyrme and WSS
models, in the case of two quark flavors and in the large-$N_c$ limit,
hence it is safe to neglect the $\eta$ mass coming from the axial anomaly.
We are aware that for a more realistic prediction we would need to include 
the anomaly term and also add the strange quark with its corresponding
mass. Such an extension to a more physically relevant analysis is left
for future work.

It would be interesting to use our result in more complex cases, for example
for multi-Skyrmion states, but we leave such a study for the future.

In the case of the Skyrme model, it was possible to perform a full
numerical computation of the moments of inertia for any value
of the pion/quark mass. This enabled us to confirm the leading-order results in
the pion mass for both the moment of inertia and for the splitting of
the Deltas.
The corresponding full numerical computation in the WSS model would
be much harder to carry out and thus we limit our analysis to the
first-order approximation linear in the pion mass.

The final result in the WSS model is a hierarchy of masses for the
Delta multiplet
$M_{\Delta^{++}} < M_{\Delta^+} < M_{\Delta^0} <M_{\Delta^-}$ and a
hierarchy of splittings
$M_{\Delta^+} - M_{\Delta^{++}}<M_{\Delta^0} - 	M_{\Delta^{+}}<M_{\Delta^-} - M_{\Delta^{0}}$.
Clearly this is valid only in the limits considered, i.e.~large-$N_c$,
large-$\lambda$ and small $m$, extrapolated to the calibrated
parameters. We want to point out that the final result is very sensitive to details. The very
fact that the Skyrmion is prolate and not oblate may change once the
$\eta$ mass coming from the axial anomaly is considered or when the
full numerical computation is done.
Moreover, the precise hierarchy of masses is also very dependent on
the precise relation between the splittings \eqref{CSsplitting} and
\eqref{prolatesplitting} which happen to be very similar in
magnitude.

\section*{Acknowledgments}

The work of L.~B.~is supported by the National Natural Science
Foundation of China (Grant No.~12150410316). 
The work of S.~B.~is supported by the INFN special research project grant
``GAST''  (Gauge and String Theories).
S.~B.~G.~thanks the Outstanding Talent Program of Henan University and
the Ministry of Education of Henan Province for partial support.
The work of S.~B.~G.~is supported by the National Natural Science
Foundation of China (Grants No.~11675223 and No.~12071111) and by the
Ministry of Science and Technology of China (Grant No.~G2022026021L).

\bibliographystyle{JHEP}
\bibliography{references}

\providecommand{\href}[2]{#2}\begingroup\raggedright\begin{thebibliography}{10}

\bibitem{Skyrme:1961vq}
T.~H.~R. Skyrme, {\it {A Nonlinear field theory}},  {\em Proc. Roy. Soc. Lond.
  A} {\bf 260} (1961) 127--138.

\bibitem{Skyrme:1962vh}
T.~H.~R. Skyrme, {\it {A Unified Field Theory of Mesons and Baryons}},  {\em
  Nucl. Phys.} {\bf 31} (1962) 556--569.

\bibitem{Zahed:1986qz}
I.~Zahed and G.~E. Brown, {\it {The Skyrme Model}},  {\em Phys. Rept.} {\bf
  142} (1986) 1--102.

\bibitem{Ma:2016npf}
Y.-L. Ma and M.~Harada, {\it {Lecture notes on the Skyrme model}},
  \href{http://arxiv.org/abs/1604.04850}{{\tt arXiv:1604.04850}}.

\bibitem{Adkins:1983ya}
G.~S. Adkins, C.~R. Nappi, and E.~Witten, {\it {Static Properties of Nucleons
  in the Skyrme Model}},  {\em Nucl. Phys. B} {\bf 228} (1983) 552.

\bibitem{Simic:1985gv}
P.~Simic, {\it {{QCD} at Large $N_c$: Skyrme or the Bag?}},  {\em Phys. Rev.
  Lett.} {\bf 55} (1985) 40.

\bibitem{Adkins:1983hy}
G.~S. Adkins and C.~R. Nappi, {\it {The Skyrme Model with Pion Masses}},  {\em
  Nucl. Phys. B} {\bf 233} (1984) 109--115.

\bibitem{Battye:2004rw}
R.~Battye and P.~Sutcliffe, {\it {Skyrmions and the pion mass}},  {\em Nucl.
  Phys. B} {\bf 705} (2005) 384--400,
  [\href{http://arxiv.org/abs/hep-ph/0410157}{{\tt hep-ph/0410157}}].

\bibitem{Battye:2006tb}
R.~Battye and P.~Sutcliffe, {\it {Skyrmions with massive pions}},  {\em Phys.
  Rev. C} {\bf 73} (2006) 055205,
  [\href{http://arxiv.org/abs/hep-th/0602220}{{\tt hep-th/0602220}}].

\bibitem{Houghton:2006ti}
C.~Houghton and S.~Magee, {\it {The Effect of pion mass on skyrme
  configurations}},  {\em EPL} {\bf 77} (2007), no.~1 11001,
  [\href{http://arxiv.org/abs/hep-th/0602227}{{\tt hep-th/0602227}}].

\bibitem{Battye:2006na}
R.~Battye, N.~S. Manton, and P.~Sutcliffe, {\it {Skyrmions and the
  alpha-particle model of nuclei}},  {\em Proc. Roy. Soc. Lond. A} {\bf 463}
  (2007) 261--279, [\href{http://arxiv.org/abs/hep-th/0605284}{{\tt
  hep-th/0605284}}].

\bibitem{Fujiwara:1984pk}
T.~Fujiwara, Y.~Igarashi, A.~Kobayashi, H.~Otsu, T.~Sato, and S.~Sawada, {\it
  {An Effective Lagrangian for Pions, $\rho$ Mesons and Skyrmions}},  {\em
  Prog. Theor. Phys.} {\bf 74} (1985) 128.

\bibitem{Meissner:1988iv}
U.~G. Meissner, N.~Kaiser, H.~Weigel, and J.~Schechter, {\it {Realistic
  Pseudoscalar - Vector Lagrangian. 2. Static and Dynamical Baryon
  Properties}},  {\em Phys. Rev. D} {\bf 39} (1989) 1956--1972. [Erratum:
  Phys.Rev.D 40, 262 (1989)].

\bibitem{Meissner:1986ka}
U.~G. Meissner, N.~Kaiser, A.~Wirzba, and W.~Weise, {\it {Skyrmions With $\rho$
  and $\omega$ Mesons as Dynamical Gauge Bosons}},  {\em Phys. Rev. Lett.} {\bf
  57} (1986) 1676.

\bibitem{Jain:1987sz}
P.~Jain, R.~Johnson, U.~G. Meissner, N.~W. Park, and J.~Schechter, {\it
  {Realistic Pseudoscalar Vector Chiral Lagrangian and Its Soliton
  Excitations}},  {\em Phys. Rev. D} {\bf 37} (1988) 3252.

\bibitem{Schechter:1999hg}
J.~Schechter and H.~Weigel, {\it {The Skyrme model for baryons}},
  \href{http://arxiv.org/abs/hep-ph/9907554}{{\tt hep-ph/9907554}}.

\bibitem{Kaymakcalan:1984bz}
O.~Kaymakcalan and J.~Schechter, {\it {Chiral Lagrangian of Pseudoscalars and
  Vectors}},  {\em Phys. Rev. D} {\bf 31} (1985) 1109.

\bibitem{Naya:2018mpt}
C.~Naya and P.~Sutcliffe, {\it {Skyrmions in models with pions and rho
  mesons}},  {\em JHEP} {\bf 05} (2018) 174,
  [\href{http://arxiv.org/abs/1803.06098}{{\tt arXiv:1803.06098}}].

\bibitem{Naya:2018kyi}
C.~Naya and P.~Sutcliffe, {\it {Skyrmions and clustering in light nuclei}},
  {\em Phys. Rev. Lett.} {\bf 121} (2018), no.~23 232002,
  [\href{http://arxiv.org/abs/1811.02064}{{\tt arXiv:1811.02064}}].

\bibitem{Witten:1979kh}
E.~Witten, {\it {Baryons in the 1/n Expansion}},  {\em Nucl. Phys. B} {\bf 160}
  (1979) 57--115.

\bibitem{Durgut:1985mu}
M.~Durgut and N.~K. Pak, {\it {Neutron - Proton Mass Difference in the Skyrme
  Model}},  {\em Phys. Lett. B} {\bf 159} (1985) 357. [Erratum: Phys.Lett.B
  162, 405 (1985)].

\bibitem{Deshpande:1976vn}
N.~G. Deshpande, D.~A. Dicus, K.~Johnson, and V.~L. Teplitz, {\it {Hadron
  Electromagnetic Mass Differences}},  {\em Phys. Rev. D} {\bf 15} (1977) 1885.

\bibitem{Ebrahim:1987mu}
A.~Ebrahim and M.~Savci, {\it {Electromagnetic Neutron Proton Mass Difference
  in the Skyrme Model}},  {\em Phys. Lett. B} {\bf 189} (1987) 343--346.

\bibitem{Epele:1988ak}
L.~N. Epele, H.~Fanchiotti, C.~A. Garcia~Canal, and R.~Mendez~Galain, {\it
  {$\rho - \omega$ Mixing and the $n - p$ Mass Difference}},  {\em Phys. Rev.
  D} {\bf 39} (1989) 1473.

\bibitem{Jain:1989kn}
P.~Jain, R.~Johnson, N.~W. Park, J.~Schechter, and H.~Weigel, {\it {The Neutron
  - Proton Mass Splitting Puzzle in Skyrme and Chiral Quark Models}},  {\em
  Phys. Rev. D} {\bf 40} (1989) 855.

\bibitem{Park:1989ak}
B.~Y. Park and M.~Rho, {\it {Neutron Proton Mass Difference in the Chiral
  Hyperbag}},  {\em Phys. Lett. B} {\bf 220} (1989) 7--13.

\bibitem{Speight:2018zgc}
J.~M. Speight, {\it {A simple mass-splitting mechanism in the Skyrme model}},
  {\em Phys. Lett. B} {\bf 781} (2018) 455--458,
  [\href{http://arxiv.org/abs/1803.11216}{{\tt arXiv:1803.11216}}].

\bibitem{Witten:1998qj}
E.~Witten, {\it {Anti-de Sitter space and holography}},  {\em Adv. Theor. Math.
  Phys.} {\bf 2} (1998) 253--291,
  [\href{http://arxiv.org/abs/hep-th/9802150}{{\tt hep-th/9802150}}].

\bibitem{Sakai:2004cn}
T.~Sakai and S.~Sugimoto, {\it {Low energy hadron physics in holographic QCD}},
   {\em Prog. Theor. Phys.} {\bf 113} (2005) 843--882,
  [\href{http://arxiv.org/abs/hep-th/0412141}{{\tt hep-th/0412141}}].

\bibitem{Sakai:2005yt}
T.~Sakai and S.~Sugimoto, {\it {More on a holographic dual of QCD}},  {\em
  Prog. Theor. Phys.} {\bf 114} (2005) 1083--1118,
  [\href{http://arxiv.org/abs/hep-th/0507073}{{\tt hep-th/0507073}}].

\bibitem{Bigazzi:2018cpg}
F.~Bigazzi and P.~Niro, {\it {Neutron-proton mass difference from gauge/gravity
  duality}},  {\em Phys. Rev. D} {\bf 98} (2018), no.~4 046004,
  [\href{http://arxiv.org/abs/1803.05202}{{\tt arXiv:1803.05202}}].

\bibitem{Wu:1990ma}
G.-W. Wu, M.-L. Yan, and K.-F. Liu, {\it {Flavor symmetry and mass splitting
  formulae for sub SU(3) skyrmion in SU(N)}},  {\em Phys. Rev. D} {\bf 43}
  (1991) 185--195.

\bibitem{Kopeliovich:1997pq}
V.~B. Kopeliovich, {\it {Semiclassical quantization of SU(3) skyrmions}},  {\em
  J. Exp. Theor. Phys.} {\bf 85} (1997) 1060--1069,
  [\href{http://arxiv.org/abs/hep-th/9707067}{{\tt hep-th/9707067}}].

\bibitem{Derrick:1964ww}
G.~H. Derrick, {\it {Comments on nonlinear wave equations as models for
  elementary particles}},  {\em J. Math. Phys.} {\bf 5} (1964) 1252--1254.

\bibitem{Bolognesi:2014ova}
S.~Bolognesi and W.~Zakrzewski, {\it {Baby Skyrme Model, Near-BPS
  Approximations and Supersymmetric Extensions}},  {\em Phys. Rev. D} {\bf 91}
  (2015), no.~4 045034, [\href{http://arxiv.org/abs/1407.3140}{{\tt
  arXiv:1407.3140}}].

\bibitem{Foster:2015cpa}
D.~Foster and N.~S. Manton, {\it {Scattering of Nucleons in the Classical
  Skyrme Model}},  {\em Nucl. Phys. B} {\bf 899} (2015) 513--526,
  [\href{http://arxiv.org/abs/1505.06843}{{\tt arXiv:1505.06843}}].

\bibitem{Gudnason:2020arj}
S.~B. Gudnason and J.~M. Speight, {\it {Realistic classical binding energies in
  the $\omega$-Skyrme model}},  {\em JHEP} {\bf 07} (2020) 184,
  [\href{http://arxiv.org/abs/2004.12862}{{\tt arXiv:2004.12862}}].

\bibitem{Hata:2007mb}
H.~Hata, T.~Sakai, S.~Sugimoto, and S.~Yamato, {\it {Baryons from instantons in
  holographic QCD}},  {\em Prog. Theor. Phys.} {\bf 117} (2007) 1157,
  [\href{http://arxiv.org/abs/hep-th/0701280}{{\tt hep-th/0701280}}].

\bibitem{Hashimoto:2008zw}
K.~Hashimoto, T.~Sakai, and S.~Sugimoto, {\it {Holographic Baryons: Static
  Properties and Form Factors from Gauge/String Duality}},  {\em Prog. Theor.
  Phys.} {\bf 120} (2008) 1093--1137,
  [\href{http://arxiv.org/abs/0806.3122}{{\tt arXiv:0806.3122}}].

\bibitem{Aharony:2008an}
O.~Aharony and D.~Kutasov, {\it {Holographic Duals of Long Open Strings}},
  {\em Phys. Rev. D} {\bf 78} (2008) 026005,
  [\href{http://arxiv.org/abs/0803.3547}{{\tt arXiv:0803.3547}}].

\bibitem{Hong:2007kx}
D.~K. Hong, M.~Rho, H.-U. Yee, and P.~Yi, {\it {Chiral Dynamics of Baryons from
  String Theory}},  {\em Phys. Rev. D} {\bf 76} (2007) 061901,
  [\href{http://arxiv.org/abs/hep-th/0701276}{{\tt hep-th/0701276}}].

\bibitem{Bolognesi:2013nja}
S.~Bolognesi and P.~Sutcliffe, {\it {The Sakai-Sugimoto soliton}},  {\em JHEP}
  {\bf 01} (2014) 078, [\href{http://arxiv.org/abs/1309.1396}{{\tt
  arXiv:1309.1396}}].

\bibitem{Hashimoto:2009hj}
K.~Hashimoto, T.~Hirayama, and D.~K. Hong, {\it {Quark Mass Dependence of
  Hadron Spectrum in Holographic QCD}},  {\em Phys. Rev. D} {\bf 81} (2010)
  045016, [\href{http://arxiv.org/abs/0906.0402}{{\tt arXiv:0906.0402}}].

\bibitem{Hashimoto:2009st}
K.~Hashimoto, N.~Iizuka, T.~Ishii, and D.~Kadoh, {\it {Three-flavor quark mass
  dependence of baryon spectra in holographic QCD}},  {\em Phys. Lett. B} {\bf
  691} (2010) 65--71, [\href{http://arxiv.org/abs/0910.1179}{{\tt
  arXiv:0910.1179}}].

\bibitem{Filewood:2002hr}
G.~R. Filewood, {\it {Delta baryons in the separation geometry model}},
  \href{http://arxiv.org/abs/hep-ph/0212328}{{\tt hep-ph/0212328}}.

\bibitem{ParticleDataGroup:2020ssz}
{\bf Particle Data Group} Collaboration, P.~A. Zyla et~al., {\it {Review of
  Particle Physics}},  {\em PTEP} {\bf 2020} (2020), no.~8 083C01.

\bibitem{Pedroni:1978it}
E.~e.~a. Pedroni, {\it {A Study of Charge Independence and Symmetry from pi+
  and pi- Total Cross-Sections on Hydrogen and Deuterium Near the 3,3
  Resonance}},  {\em Nucl. Phys. A} {\bf 300} (1978) 321--347.

\bibitem{Abaev:1995cx}
V.~V. Abaev and S.~P. Kruglov, {\it {Phase shift analysis of pi p scattering in
  the energy region from 160-MeV to 600-MeV}},  {\em Z. Phys. A} {\bf 352}
  (1995) 85--96.

\bibitem{Bernicha:1995gg}
A.~Bernicha, G.~Lopez~Castro, and J.~Pestieau, {\it {Pion - proton scattering
  and isospin breaking in the Delta0 - Delta++ system}},  {\em Nucl. Phys. A}
  {\bf 597} (1996) 623--635, [\href{http://arxiv.org/abs/hep-ph/9508388}{{\tt
  hep-ph/9508388}}].

\bibitem{Gridnev:2004mk}
A.~B. Gridnev, I.~Horn, W.~J. Briscoe, and I.~I. Strakovsky, {\it {The K-matrix
  approach to the Delta - resonance mass splitting and isospin violation in
  low-energy pi-N scattering}},  {\em Phys. Atom. Nucl.} {\bf 69} (2006)
  1542--1551, [\href{http://arxiv.org/abs/hep-ph/0408192}{{\tt
  hep-ph/0408192}}].

\end{thebibliography}\endgroup

\end{document}